\documentclass [12pt]{article}
\usepackage{amssymb}
\usepackage{amsmath}
\usepackage{amsfonts}
\usepackage{bm}
\usepackage{graphicx}
\begin{document}

\begin{center}

\Large{\bf{A generating functional approach to the Hubbard model}}

\normalsize

\ \\
\ \\
Yu.A. Izyumov, N.I. Chaschin, D.S. Alexeev

\ \\
\em{Institute for Metal Physics, Ural Division of the Russian
Academy of Sciences, \\ 620219, Ekaterinburg, S.Kovalevskaya
str.18, Russia}

\normalfont

\vspace{1.0cm} F. Mancini

\ \\
\em{Dipartimento di Fisica ``E.R. Caianiello", Universit\`{a}
degli Studi di Salerno, \\ Via S. Allende, I-84081 Baronissi (SA),
Italy}

\end{center}


\ \\

\begin{abstract}
\noindent The method of generating functional, suggested for
conventional systems by Kadanoff and Baym, is generalized to the
case of strongly correlated systems, described by the Hubbard $X$
operators. The method has been applied to the Hubbard model with
arbitrary value $U$ of the Coulomb on-site interaction. For the
electronic Green's function $\mathcal{G}$ constructed for
Fermi-like $X$ operators, an equation using variational
derivatives with respect to the fluctuating fields has been
derived and its multiplicative form has been determined. The
Green's function is characterized by two quantities: the self
energy $\Sigma$ and the terminal part $\Lambda$. For them we have
derived the equation using variational derivatives, whose
iterations generate the perturbation theory near the atomic limit.
Corrections for the electronic self-energy $\Sigma$ are calculated
up to the second order with respect to the parameter $W/U$ ($W$
width of the band), and a mean field type approximation was
formulated, including both charge and spin static fluctuations.
This approximation is actually equivalent to the one used in the
method of Composite Operators, and it describes an insulator-metal
phase transition at half filling reasonably well.

The equations for the Bose-like Green's functions have been
derived, describing the collective modes: the magnons and
doublons. The main term in this equation represents variational
derivatives of the electronic Green's function with respect to the
corresponding fluctuating fields. The properties of the poles of
the doublon Green's functions depend on electronic filling. The
investigation of the special case $n=1$ demonstrates that the
doublon Green's function has a soft mode at the wave vector
$\bm{Q}=(\pi,\pi,\dots)$, indicating possible instability of the
uniform paramagnetic phase relatively to the two sublattices
charge ordering. However this instability should compete with an
instability to antiferromagnetic ordering.

The generating functional method with the $X$ operators could be
extended to the other models of strongly correlated systems.
\end{abstract}


\section{Introduction}
The Hubbard model is one of the basic models in the theory of
strongly correlated systems. During its forty years of lifetime
numerous approaches have been proposed for the investigation of
the possible states of the system, the spectrum of its
quasi-particles and the collective modes, the transport
properties, the different types of ordered states and the phase
transitions among them. Such long period of development of a model
which could look simple at a first glance --- since it contains
only two parameters, the bare bandwidth $W$ and the on-site
Coulomb repulsion $U$ --- is determined by the circumstance that
the case $U \geq W$ is of main physical significance. But just in
this case the theory does not contain a small parameter. Already
the first researchers tried to avoid perturbative theories and
used different non-perturbative approaches. Starting from the
pioneering works of Hubbard \cite{hu1} \cite{hu2} \cite{hu3}, the
method of decoupling of the double-time Green's functions (GF) was
treated successfully. The works based on projecting the equations
of motion for the basic operators come here \cite{ro} \cite{go}
\cite{pla}. The most productive application of this approach has
been done with the method of composite operators \cite{man}
\cite{ave1} \cite{ave2} \cite{man2} used widely not only for the
Hubbard model but also for many other models \cite{ish} of
strongly correlated electronic systems. The method of the spectral
density moments uses in essence the cut short of the equations of
motions for the basic operators as well \cite{no} \cite{her}. Also
the variational method of Gutzwiller belongs to the
non-perturbative approaches \cite{gutz}, and made it possible to
investigate qualitatively the behavior of a vast class of strongly
correlated systems during the last four decades. The method of
slave particles (slave bosons) represents an important direction
of investigation also \cite{bar}\cite{kot}\cite{lil}\cite{fre}.
The basic operators are expressed through a product of
conventional Fermi and Bose operators with subsequent exclusion of
unphysical states. The suitable choice of a slave particle
representation makes it possible to catch the physics of low
energy states in the scope of the mean field approximation.
Unfortunately there is no standard recipe for constructing such
representations, and it is not always clear which one among the
possible representations is the most adequate.

During the last decade the method of the dynamical mean field
theory (DMFT) has become quite popular \cite{metz}\cite{geo1}.
By means of this
method it has been possible to investigate the behavior of almost
all the models in the theory of strongly correlated systems in the
region of strong and intermediate interactions. Apparently DMFT is
the most efficient method of investigation of these systems,
although not exempt from some defects: it demands a huge amount of
computations and has problems with the description of collective
modes (see the review \cite{geo2}). We do not mention here the numerical
methods like Quantum Monte Carlo and small cluster
diagonalization, because we concentrate our efforts on the
analytical approaches.

We want to pay attention to one of the analytical approaches where
there is a possibility to derive a consistent perturbative theory
with respect to the parameter $W/U$. Definitely, such an approach
corresponds to the perturbative theory near the atomic limit. The
approach is based on the introduction of a generating functional
$Z[V]$, describing the interaction of the system with fluctuating
fields depending on space and time. This functional corresponds to
the generalization of the partition function of the system for the
case of interactions with external fluctuating fields. For a
proper choice of the $V$ operator the different GFs of the system
are expressed through variational derivatives with respect to
fluctuating fields.

At the beginning this method was developed for a weak interaction
by Kadanoff and Baym \cite{ba}\cite{ka} forty years ago. It could be
generalized to a strongly correlated system when we express the
Hamiltonian through some basic operators taking into account the
correlations (for instance the Hubbard $X$ operators)[3] instead
of the conventional ones. The first time such an approach has been
applied to the Hubbard model was in the limit $U\to\infty$ (with
an additional small parameter $1/N$, where $N$ is the degeneracy
of the electronic states) \cite{ru}. Afterwards this approach has been
developed farther in the works \cite{ku}\cite{ze}\cite{iz1}\cite{iz2}.

Recently we have provided a general framework for the generating
functional approach (GFA) and we have applied it to a set of basic
models of spin and strongly correlated electronic systems:
Heisenberg model, Hubbard model for $U\rightarrow\infty$,
$tJ$-model, $sd$-model, double exchange model \cite{iz1}\cite{iz2}. The results
of these investigations have been generalized in the monograph
\cite{iz3}, published in Russian, and in the course of lectures
delivered in an international school \cite{iz4}.

In this paper we apply the GFA to the Hubbard model with a finite
Coulomb interaction $U$. Supposing that $U$ is large but of the
order of $W$ we express the Hamiltonian of the model in terms of
the $X$ operators and calculate the electronic and bosonic GFs.
The latter describes the two types of collective modes: magnons
and doublons.

The electronic GF is a matrix with respect to the spin index
$\sigma$, the index $\alpha$, indicating the Hubbard subbands, and
the index $\nu$ corresponding to the particle-hole representation.
We have derived the equation in the variational derivatives with
respect to fluctuating fields for it. Because the basic operators
do not commute on $c$-values, the electronic Green's function is
characterized by two functions of four-momenta: the self-energy
$\Sigma$ and the terminal part $\Lambda$. For $\Sigma$ and
$\Lambda$ the equations with the variational derivatives have been
derived too, whereas it is possible to make iterations with
respect to the parameter $W/U$. Just these iterative series
represent the perturbative theory near the atomic limit
\cite{har}. We have limited ourselves to the first and second
order corrections for $\Sigma$ and extracted from them a mean
field type $\Sigma_{MF}$ part, which includes contributions
depending only on the wave vector $\bm{k}$, but not on the
frequency. $\Sigma_{MF}$ consists of a term giving a shift to the
Hubbard subbands and renormalizing its width. The last term was
extracted from the second order correction $\Sigma_{2}^{\prime}$,
which is an ``uncutable" term (with respect to the hopping matrix
element), while a ``cutable" term $\Sigma_{2}$ describes the
dynamical interaction with boson-type excitations. A procedure of
extraction of the static part from $\Sigma_{2}^{\prime}$ was
borrowed from the Composite Operator Method (COM)
\cite{man}\cite{ave1}\cite{ave2}\cite{man2}. The main idea of this
approach is that bosonic correlators, describing for example
static fluctuations of charge, spin and pair, should not be
calculated by some uncontrollable approximation (like decoupling
or use of the equation of motion), but must be determined by means
of general properties of the electronic GF [10].

The GFA, restricted to the mean field approximation, and the COM,
restricted to a two-pole approximation, have a different structure
for the electronic GF. In spite of this, the results obtained by these
two methods for
different properties of the Hubbard model turned out to be in very
good agreement. In particular, such mean field GFs give two
quasiparticle subbands with a gap between them, which vanishes for
half-filling at some critical value $U=U_{c}$, and an
insulator-metal phase transition occurs. Detailed comparison of
the mean field approximation in GFA and COM will be discussed
below.

Using the electron GFs we found, we can calculate Bose-like GFs
for plasmons, magnons and doublons. In this paper we study only
doublons -- collective modes, describing motion of double occupied
states of the lattice sites. The equation for the doublon GF has
been derived. This equation contains variational derivatives of
the electronic GF with respect to the corresponding fluctuating
fields, coupled with charge densities. In the mean field
approximation for the electronic $\Sigma$ we have obtained the
closed equation for the doublon GF. For the paramagnetic state at
half filling ($n=1$) the doublon GF has a soft mode at momentum
$\bm{Q}=\bm{\pi}=(\pi,\pi,\dots)$. It indicates a possible
instability of the uniform state against a charge density wave
formation. When the filling deviates from unity ($n<1$), the pole
of the doublon GF has a gap $U-2\mu$, thus having the activative
character.

The content of the paper is the following. In part 2, based on the
$X$ operators formalism, the GFA is constructed. In part 3 it is
derived the equation of motion for the electronic GF in the form
of equation with variational derivatives. This equation is
decoupled into two: one for the self energy and one for the
terminal part. In part 4 the iterations of these equations with
respect to the parameter $W/U$ are implemented and the GF in the
``Hartree-Fock approximation" is calculated. In part 5 we
formulate a mean field approximation and compare GFA and COM
approaches. A Bose-like GF for doublons is calculated in part 6
with the electronic GF taken in the mean field approximation. In
part 7 we calculate the doublon susceptibility in the
hydrodynamical regime. Finally in part 8 we discuss the obtained
results and propose suggestions for further study of the Hubbard
model.


\section{Introduction of the generating functional}

Let us consider the conventional Hubbard model for nondegenerate
states. In terms of the Fermi operators the model Hamiltonian is
\begin{equation}
\mathcal{H}=\sum\limits_{ij\sigma}t_{ij}
c_{i\sigma}^{\dag}c_{j\sigma} +
U\sum\limits_{i} n_{i\uparrow}n_{i\downarrow} ,
\label{eq:2.1}
\end{equation}
where $c_{i\sigma}(c_{i\sigma}^\dag)$ is the operator of annihilation
(creation) of an electron on the site $i$ with spin $\sigma$,
$n_{i\sigma}=c_{i\sigma}^{\dag}c_{i\sigma}$ is the electron number on
the same site with the given spin. Under the condition of a strong
on-site Coulomb repulsion $U>zt$ (where $t$ is the hopping matrix
element for the nearest neighbors and $z$ is the coordination number)
it is useful to express the Hamiltonian (\ref{eq:2.1}) in terms of
the $X$ operators. The operator $X^{pq}_i$ for the site $i$
describes the transitions between the four possible states
$p=|0\rangle, |\sigma\rangle, |\bar{\sigma}\rangle, |2\rangle$
--- without any electron, with one electron possessing the spin
projection $\sigma$ or $-\sigma$ and a pair of electrons,
respectively.

The $X$ operators could be represented through the conventional
Fermi operators by means of the relations
\begin{alignat}{3}
&X_{i}^{\sigma 0}=c_{i\sigma}^\dag (1-n_{i\bar{\sigma}}),&{\qquad}
&X_{i}^{2 \sigma}=\sigma c_{i\bar{\sigma}}^\dag n_{i\sigma},\nonumber \\
&X_i^{\sigma \bar{\sigma}}=c_{i\sigma}^\dag c_{i\bar{\sigma}},&{\qquad}
&X_i^{20}=\sigma c_{i\bar{\sigma}}^\dag c_{i\sigma}^\dag, \\
&X_i^{\sigma \sigma}=n_{i\sigma}(1-n_{i\bar{\sigma}}),&{\qquad}
&X_i^{22}=n_{i\sigma}n_{i\bar{\sigma}},\nonumber\\
&X_i^{00}=(1-n_{i\sigma})(1-n_{i\bar{\sigma}})\nonumber.
\label{eq:2.2}
\end{alignat}
The operators $X_i^{\sigma 0}$ and $X_i^{2\sigma}$ describe the
correlated creation of an electron and are Fermi-like
$f$-operators; $X_i^{\sigma \bar{\sigma}}$ and $X_i^{20}$ describe
the flip of a spin on a site and the creation of a pair; they are
Bose-like $b$-operators, respectively. The remaining $X$'s are
called diagonal. We note that there are the hermitian-conjugate
operators $(X_i^{pq})^\dag=X_i^{qp}$. The sixteen $X$ operators
comprise thus the whole set, forming the algebra with the
corresponding property of the product
\begin{equation}
X_i^{rs}X_i^{pq}=\delta_{sp}X_i^{rq}.
\label{eq:2.3}
\end{equation}
and the permutation relations of the anticommuting type for the
$f$-operators while commutating for the $b$-operators. We note that
the conventional Fermi operators are expressed through the linear
combinations of the $X$ operators of the $f$-type
\begin{equation}
c^{\dag}_{i\sigma}=X^{\sigma 0}_{i}-\sigma X^{2\bar{\sigma}}_{i},\qquad
c_{i\sigma}=X^{0\sigma}_{i}-\sigma X^{\bar{\sigma}2}_{i}.
\label{eq:2.4}
\end{equation}
These relations express the motion of the correlated electrons in the two
Hubbard subbands.

It is convenient to introduce the two-component spinors for the
the $f$-operators:
\begin{alignat}{2}
\varPsi(i\sigma)&=
\begin{pmatrix}
X_{i}^{0\sigma} \\
\bar{\sigma}X^{\bar{\sigma}2}_{i}
\end{pmatrix},{\qquad}
&\varPsi^{\dag}(i\sigma)=\left(X_{i}^{\sigma0},\,
\bar{\sigma} X_{i}^{2\bar{\sigma}}\right).
\label{eq:2.5}
\end{alignat}
Then the Hamiltonian (\ref{eq:2.1}) is represented as
$\mathcal{H}=\mathcal{H}_{0}+\mathcal{H}_{1}$, where
\begin{equation}
\mathcal{H}_{0}=\sum_{i}\left(\sum_{\sigma}
\varepsilon_{\sigma}X_{i}^{\sigma\sigma}+
\varepsilon_{2}X_{i}^{22}\right),
\label{eq:2.6}
\end{equation}
\begin{equation}
\mathcal{H}_{1}=\sum_{ij}\sum_{\sigma}\sum_{\alpha_{1}\alpha_{2}}
\varPsi^{\dag}_{\alpha_{1}}(i\sigma)
t_{\alpha_{1}\alpha_{2}}(ij)
\varPsi_{\alpha_{2}}(j\sigma).
\label{eq:2.7}
\end{equation}
Here we added to Hamiltonian (2.1) the term 
$\displaystyle\sum\limits_{i\sigma}(-\mu -\sigma
\frac{\displaystyle h}{\displaystyle2})n_{i \sigma}$, where $\mu$ is the chemical potential 
and $\displaystyle h$ is the external magnetic field, that is why new notation appears:
$\varepsilon_{\sigma}=
-\sigma\frac{\displaystyle h}{\displaystyle2}-\mu,{\quad}
\varepsilon_{2}=U-2\mu$. In the quadratic form
(\ref{eq:2.7}) $\varPsi_{\alpha}(i\sigma)$ represents the
component of the spinor $\varPsi(i\sigma),\, (\alpha=1,2)$; in
addition we have introduced the matrix
\begin{alignat}{2}
t_{\alpha\beta}(ij)=t_{ij}\Im_{\alpha\beta},
&{\qquad}\Im&=\begin{pmatrix}
1 & 1 \\
1 & 1
\end{pmatrix}.
\label{eq:2.8}
\end{alignat}
Note that the index $\alpha$ numerates the Hubbard subbands.
With the help of the rule of multiplication (\ref{eq:2.3}) for $X$
operators, one can write the permutation relations of the spinor
$f$-operators:
\begin{equation}
\left.
\begin{split}
&\left[
\varPsi(i\sigma)\otimes\varPsi^{\dag}(j\sigma)
\right]_{+}=
\delta_{ij}F_{i}^{\sigma}\\
&\left[
\varPsi(i\sigma)\otimes\varPsi^{\dag}(j\bar{\sigma})
\right]_{+}=
\delta_{ij}X_{i}^{\bar{\sigma}\sigma}\tau^{z}\\
&\left[
\varPsi(i\sigma)\otimes\varPsi(j\bar{\sigma})
\right]_{+}=
\delta_{ij}\sigma X_{i}^{02}(i \tau^{y})
\end{split}
\right\},
\label{eq:2.9}
\end{equation}
where $\tau^{x}, \tau^{y}, \tau^{z}$ are the Pauli matrices, and
$F_{i}^{\sigma}$ is a $2\times2$ matrix, composed of $X$
operators:
\begin{equation}
F_{i}^{\sigma}=
\begin{pmatrix}
X_{i}^{00}+X_{i}^{\sigma\sigma} & 0 \\
0 & X_{i}^{\bar{\sigma}\bar{\sigma}}+X_{i}^{22}
\end{pmatrix}.
\label{eq:2.10}
\end{equation}

The permutation relations between $f$- and $b$-operators have a
commutator character:
\begin{equation}
\left.
\begin{split}
&\left[
\varPsi(i\sigma_{1}),X_{j}^{\sigma_{2}\bar{\sigma}_{2}}
\right]_{-}=
\delta_{ij}\delta_{\sigma_{1}\sigma_{2}}
\varPsi(i\bar{\sigma_{1}})\\
&\left[
\varPsi(i\sigma_{1}),X_{j}^{20}
\right]_{-}=
\delta_{ij}\bar{\sigma}_{1}
\varPsi^{\dag}(i\bar{\sigma}_{1})\tau^{x}\\
\end{split}
\right\}.
\label{eq:2.11}
\end{equation}
In other cases of permutations, relations of type (\ref{eq:2.9})
and (\ref{eq:2.11}) give zero.

Thus, an anticommutator of two $\varPsi$-operators is expressed
either through a diagonal or a  $b$-operator, but the commutator
of $\varPsi$- and $b$-operators is naturally a $\varPsi$-operator.
Note the two relations
\begin{equation}
\left(
X_{i}^{pq}
\right)^{\dag}=X_{i}^{qp},
\label{eq:2.12}
\end{equation}
\begin{equation}
X_{i}^{00}+X_{i}^{\sigma\sigma}+
X_{i}^{\bar{\sigma}\bar{\sigma}}+X_{i}^{22}=1,
\label{eq:2.13}
\end{equation}
which complete the algebra of the $X$ operators.

Let us write the equation of motion for the $f$-operator. For the
thermodynamical time $\tau\quad(-\beta \leqslant \tau \leqslant
\beta, \quad \beta={1}/{kT})$ we start from the Heisenberg
equation
\begin{equation*}
\dot{\varPsi}(\scriptstyle 1 \displaystyle
\sigma)=-[\varPsi(\scriptstyle 1 \displaystyle
\sigma),\mathcal{H}]\,,
\end{equation*}
which could be written in the case of the Hamiltonian
(\ref{eq:2.6}) -- (\ref{eq:2.7}) in the following form
\begin{align}
\dot{\varPsi}(\scriptstyle 1 \displaystyle\sigma_{1})=
&-E_{1}^{\sigma_{1}} \varPsi(\scriptstyle 1
\displaystyle\sigma_{1})- F_{1}^{\sigma_{1}}\hat{t}(\scriptstyle
11^{\prime} \displaystyle) \varPsi(\scriptstyle
1^{\prime}\displaystyle \sigma_1) \label{eq:2.14}
\\
&-X_{1}^{\bar{\sigma_{1}}\sigma_{1}}\tau^{z}
\hat{t}(\scriptstyle 11^{\prime}\displaystyle )
\varPsi(\scriptstyle 1^{\prime}\displaystyle\bar{\sigma_{1}})
+\bar{\sigma}_{1}
\varPsi^{\dag}(\scriptstyle 1^{\prime}\displaystyle\bar{\sigma_{1}})
\hat{t}(\scriptstyle 1^{\prime}1\displaystyle)i\tau^{y}X_{1}^{02}.
\nonumber
\end{align}
Here a double-row matrix with respect to the spinor index was
introduced
\begin{equation}
E_{1}^{\sigma}=
\begin{pmatrix}
\varepsilon_{\sigma} & 0 \\
0 & \varepsilon_{\sigma}+U
\end{pmatrix}.
\label{eq:2.15}
\end{equation}

Here and in the following the numerical indexes indicate the
four-dimensional coordinates including the site and the time
$\tau$, i.e. $\scriptstyle 1 \displaystyle =(i_{1},\tau_{1}),...$;
a summation over the primed indexes is understood (it is a
summation over the sites $i$ and an integration over the time
$\tau$). And finally the value
\begin{equation}
\hat{t}(\scriptstyle 11^{\prime} \displaystyle)=
\delta(\tau_{1}-\tau_{1}^{\prime})t_{i_{1}i_{1^{\prime}}}
\Im\equiv t(\scriptstyle 11^{\prime} \displaystyle)
\Im,
\label{eq:2.16}
\end{equation}
has been introduced, representing the matrix over the spinor indexes (the
last circumstance has been specified by the symbol $\hat{t}\,$).

Thus the operator $\dot{\varPsi}$ represents the linear
combination of the $f$-operators, with the bosonic $b$-operators
as the coefficients, and the matrixes $E$ and $\hat{t}$ too.

Following the method we have applied many times to different
quantum models \cite{iz1}\cite{iz2}\cite{iz3}\cite{iz41}, we
introduce the generating functional
\begin{equation}
Z[V]=\text{Tr}\left(\mathrm{e}^{-\beta \mathcal{H}}T\mathrm{e}^{-V}\right)
\equiv\mathrm{e}^{\Phi},
\label{eq:2.13}
\end{equation}
where $T$ is the symbol of the chronological product and the trace
is taken over the whole set of variables of the system.

For the Hamiltonian (\ref{eq:2.6}) -- (\ref{eq:2.7}) it is
convenient to choose the operator $V$ in the form
\begin{align}
V=v_{1^{\prime}}^{00}X_{1^{\prime}}^{00}+
&v_{1^{\prime}}^{22}X_{1^{\prime}}^{22}+
v_{1^{\prime}}^{\sigma^{\prime}\sigma^{\prime}}
X_{1^{\prime}}^{\sigma^{\prime}\sigma^{\prime}}+
v_{1^{\prime}}^{\sigma^{\prime}\bar{\sigma}^{\prime}}
X_{1^{\prime}}^{\bar{\sigma}^{\prime}\sigma^{\prime}}
\label{eq:2.19}
\\
+&v_{1^{\prime}}^{02}X_{1^{\prime}}^{20}+
v_{1^{\prime}}^{20}X_{1^{\prime}}^{02}.  
\nonumber
\end{align}

It represents the linear combination of the whole diagonal and
$b$-operators with the single point fields $v$. Thus,
differentiating the equation $Z[V]$ with respect to the different
$v$'s, we can express the different GFs through the variational
derivatives with respect to the corresponding fields. For
instance, for the single particle Bose-like GFs of the plasmons,
magnons and the doublons we have the expressions:
\begin{equation}
\mathcal{N}^{\sigma_{1}\sigma_{2}}
(\scriptstyle 12 \displaystyle)=
-\langle TX_{1}^{\sigma_{1}\sigma_{1}}
X_{2}^{\sigma_{2}\sigma_{2}}\rangle_{V}=
-\frac{1}{Z[V]}\frac{\displaystyle\delta^{2}Z[V]}
{\displaystyle\delta v_{1}^{\sigma_{1}\sigma_{1}}
\delta v_{2}^{\sigma_{2}\sigma_{2}}},
\label{eq:2.20}
\end{equation}
\begin{equation}
\mathcal{D}^{\sigma\bar{\sigma}}
(\scriptstyle 12 \displaystyle)=
-\langle TX_{1}^{\sigma\bar{\sigma}}
X_{2}^{\bar{\sigma}\sigma}\rangle_{V}=
-\frac{\displaystyle\delta^{2}\Phi}
{\displaystyle\delta v_{1}^{\sigma\bar{\sigma}}
\delta v_{2}^{\bar{\sigma}\sigma}},
\label{eq:2.21}
\end{equation}
\begin{equation}
\mathcal{D}^{02}
(\scriptstyle 12 \displaystyle)=
-\langle TX_{1}^{02}X_{2}^{20}\rangle_{V}=
-\frac{\displaystyle\delta^{2}\Phi}
{\displaystyle\delta v_{1}^{02}\delta v_{2}^{20}}.
\label{eq:2.22}
\end{equation}
Here and further symbol $\langle\dots\rangle_{V}\equiv\langle\dots\mathrm{e}^{-V}\rangle$,
where $\langle\dots\rangle$ means averading over Gibbs ansamble with Hamiltonian
$\mathcal{H}$.
Having been introduced in such a way, the GFs are functionals of
the fluctuating fields. Directing these fields to zero after
taking the variational derivatives, we shall obtain the actual
GFs, describing our system. The fermionic GF cannot be obtained by
differentiation of $Z[V]$($\Phi$) with respect to the single-point fields
and it is necessary to determine the equation of motion for them.


\section{Equations of motion for electron Green's function}

We make use of the general equation of motion (see Appendix) and
write it for the expression $((T\varPsi_{1}\varPsi^{\dag}_{2}))$,
determining the electronic GF:
\begin{align}
&\frac{\partial}{\partial\tau_{1}}
(\!(T\varPsi_{\alpha_{1}}(\scriptstyle 1 \displaystyle \sigma_{1})
\varPsi^{\dag}_{\alpha_{2}}(\scriptstyle 2
\displaystyle\sigma_{2}) e^{-V})\!)=
(\!(T\{\varPsi_{\alpha_{1}}(\scriptstyle 1
\displaystyle\sigma_{1}), \varPsi^{\dag}_{\alpha_{2}}(\scriptstyle
2 \displaystyle\sigma_{2})\}_{+} e^{-V})\!) \label{eq:3.1}
\\
&+(\!(T\dot{\varPsi}_{\alpha_{1}}(\scriptstyle 1 \displaystyle\sigma_{1})
\varPsi_{\alpha_{2}}^{\dag}(\scriptstyle 2 \displaystyle\sigma_{2})
e^{-V})\!)-
(\!(T\{\varPsi_{\alpha_{1}}(\scriptstyle 1 \displaystyle\sigma_{1}),V\}_{-}
\varPsi^{\dag}_{\alpha_{2}}(\scriptstyle 2 \displaystyle\sigma_{2})
e^{-V})\!).  \nonumber
\end{align}
Let us calculate now the anticommutator and the commutator of the $\varPsi$%
-operators in (\ref{eq:3.1}). According to relations
(\ref{eq:2.9}) and (\ref{eq:2.11}), we have:
\begin{equation}
\{\varPsi_{\alpha_{1}}(\scriptstyle 1 \displaystyle\sigma_{1}),
\varPsi^{\dag}_{\alpha_{2}}(\scriptstyle 2 \displaystyle\sigma_{2})\}_{+}=
\delta_{12}
\left(\delta_{\sigma_{1}\sigma_{2}}
\left(F_{1}^{\sigma_{1}}
\right)_{\alpha_{1}\alpha_{2}}+
\delta_{\bar{\sigma_{1}}\sigma_{2}}\tau^{z}_{\alpha_{1}\alpha_{2}}
X_{1}^{\bar{\sigma_{1}}\sigma_{1}}
\right),
\label{eq:3.2}
\end{equation}
\begin{equation}
\{\varPsi(\scriptstyle 1 \displaystyle\sigma_{1}),V\}_{-}=
W_{1}^{\sigma_{1}}
\varPsi(\scriptstyle 1 \displaystyle\sigma_{1})+
v_{1}^{\bar{\sigma_{1}}\sigma_{1}}
\varPsi(\scriptstyle 1 \displaystyle\bar{\sigma_{1}})+
\bar{\sigma}_{1}v_{1}^{02}
\varPsi^{\dag}(\scriptstyle 1 \displaystyle\bar{\sigma_{1}})
\tau^{x}.
\label{eq:3.3}
\end{equation}
Here $W$ is the double-row matrix composed with the fluctuating
fields:
\begin{equation}
W_{1}^{\sigma}=
\begin{pmatrix}
v_{1}^{\sigma\sigma}-v_{1}^{00} & 0 \\
0 & v_{1}^{22}-v_{1}^{\bar{\sigma}\bar{\sigma}}
\end{pmatrix}.
\label{eq:3.4}
\end{equation}
After the substitution of expression (\ref{eq:2.14}) and the
commutators in equation (\ref{eq:3.1}), the latter could be represented in
the form:
\begin{align}
G_{0V}^{-1}
(\scriptstyle 1 \displaystyle\sigma_{1},\,
\scriptstyle 1^{\prime} \displaystyle\sigma_{1}^{\prime})
&(\!(T\varPsi(\scriptstyle 1^{\prime} \displaystyle\sigma_{1}^{\prime})
\varPsi^{\dag}(\scriptstyle 2 \displaystyle\sigma_{2})e^{-V})\!)=
\nonumber\\
=&-\delta_{12}\left[\delta_{\sigma_{1}\sigma_{2}}
(\!(TF_{1}^{\sigma_{1}}e^{-V})\!)+
\delta_{\bar{\sigma}_{1}\sigma_{2}}\tau^{z}
(\!(TX_{1}^{\bar{\sigma}_{1}\sigma_{1}}e^{-V})\!)\right]
\nonumber\\
&+\bar{\sigma}_{1}v_{1}^{02}\tau^{x}
(\!(T\varPsi^{\dag}(\scriptstyle 1 \displaystyle\bar{\sigma}_{1})
\varPsi^{\dag}(\scriptstyle 2 \displaystyle\sigma_{2})e^{-V})\!)
\label{eq:3.5} \\
&+(\!(TF_{1}^{\sigma_{1}} \hat{t}(\scriptstyle 11^{\prime}
\displaystyle) \varPsi(\scriptstyle 1^{\prime}
\displaystyle\sigma_{1}) \varPsi^{\dag}(\scriptstyle 2
\displaystyle\sigma_{2})e^{-V})\!)
\nonumber\\
&+\tau^{z} \hat{t}(\scriptstyle 11^{\prime} \displaystyle)
(\!(TX_{1}^{\bar{\sigma}_{1}\sigma_{1}} \varPsi(\scriptstyle
1^{\prime} \displaystyle\bar{\sigma}_{1})
\varPsi^{\dag}(\scriptstyle 2 \displaystyle\sigma_{2})e^{-V})\!)
\nonumber\\
&+\sigma_{1}(\!(TX_{1}^{02}
\varPsi^{\dag}(\scriptstyle 1^{\prime} \displaystyle\bar{\sigma}_{1})
\varPsi^{\dag}(\scriptstyle 2 \displaystyle\sigma_{2})e^{-V})\!)
\hat{t}(\scriptstyle 1^{\prime}1 \displaystyle)i\tau^{y}.  \nonumber
\end{align}
Here the quantity
\begin{equation}
G_{0V}^{-1}
(\scriptstyle 1 \displaystyle\sigma_{1},\,
\scriptstyle 2 \displaystyle\sigma_{2})=
\left\{
\left(
-\frac{\partial}{\partial\tau_{1}}-E_{1}^{\sigma_{1}}
\right)
\delta_{\sigma_{1}\sigma_{2}}-
W_{1}^{\sigma_{1}}\delta_{\sigma_{1}\sigma_{2}}-
v_{1}^{\bar{\sigma_{1}}\sigma_{1}}\tau^{0}
\delta_{\bar{\sigma_{1}} \sigma_{2}}
\right\}
\delta_{12},
\label{eq:3.6}
\end{equation}
has been introduced, which defines the zeroth-order approximation propagator of the electrons in the
fluctuating single-point fields. This quantity is the $2\times2$
matrix with respect to spinor indexes. Expressing the mixed GFs
through the variational derivatives of $Z[V]$, we can represent
the obtained equation as
\begin{align}
G_{0V}^{-1}(\scriptstyle 1 \displaystyle\sigma_{1}, &\scriptstyle
1^{\prime} \displaystyle\sigma_{1}^{\prime})
(\!(T\varPsi(\scriptstyle 1^{\prime}
\displaystyle\sigma_{1}^{\prime}) \varPsi^{\dag}(\scriptstyle 2
\displaystyle\sigma_{2})e^{-V})\!)
\nonumber\\
=&-\delta_{12}\hat{a}_{1}(\sigma_{1}\sigma_{2})Z[V]+
\hat{a}_{1}(\sigma_{1}\sigma_{1}^{\prime}) \hat{t}(\scriptstyle
11^{\prime}\displaystyle) (\!(T\varPsi(\scriptstyle
1^{\prime}\displaystyle\sigma_{1}^{\prime})
\varPsi^{\dag}(\scriptstyle 2 \displaystyle\sigma_{2})e^{-V})\!)
\label{eq:3.7}\\
&-\sigma_{1}\left[v_{1}^{02}\tau^{x}\delta_{11^{\prime}}-
i\tau^{y}\,\hat{\tilde{t}}
(\scriptstyle 11^{\prime} \displaystyle)
\frac{\displaystyle\delta}{\displaystyle\delta
v_{1}^{02}}\right]
(\!(T\varPsi^{\dag}(\scriptstyle 1^{\prime} \displaystyle\bar{\sigma}_{1})
\varPsi^{\dag}(\scriptstyle 2 \displaystyle\sigma_{2})e^{-V})\!).
\nonumber
\end{align}
Here the double-row matrix is the differential operator with
respect to the single point fluctuating fields:
\begin{equation}
\hat{a}_{1}(\sigma_{1},\sigma_{2})=
\left(\delta_{\sigma_{1}\sigma_{2}}
\hat{F}_{1}^{\sigma_{1}}-
\delta_{\bar{\sigma_{1}}\sigma_{2}}\tau^{z}
\frac{\delta}
{\delta v_{1}^{\bar{\sigma_{1}}\sigma_{1}}}\right),
\label{eq:3.8}
\end{equation}
where
\begin{equation}
\hat{F}_{1}^{\sigma}=-
\begin{pmatrix}
\frac{\displaystyle\delta}
{\displaystyle\delta v_{1}^{00}}+
\frac{\displaystyle\delta}
{\displaystyle\delta v_{1}^{\sigma\sigma}}
& 0 \\
0 & \frac{\displaystyle\delta} {\displaystyle\delta
v_{1}^{\bar{\sigma}\bar{\sigma}}}+ \frac{\displaystyle\delta}
{\displaystyle\delta v_{1}^{22}}
\end{pmatrix}.
\label{eq:3.9}
\end{equation}Also, let us note that
$\tilde{\hat{t}}$ is the transposed matrix of $\hat{t}$.

As usual, we pass from the functional $Z[V]$ to the functional $\Phi[V]$
using the substitution:
\begin{equation}
Z[V]=e^{\Phi[V]}.
\label{eq:3.10}
\end{equation}

Then, the equation (\ref{eq:3.5}) results in a direct equation for
the electronic GF:
\begin{align}
[ G_{0V}^{-1} (\scriptstyle 1 \displaystyle\sigma,\, \scriptstyle
1^{\prime} \displaystyle\sigma_{1}^{\prime})&-
(\hat{a}_{1}(\sigma_{1}\sigma_{1}^{\prime})\Phi)
\hat{t}(\scriptstyle 11^{\prime} \displaystyle)-
\hat{a}_{1}(\sigma_{1}\sigma_{1}^{\prime}) \hat{t}(\scriptstyle
11^{\prime} \displaystyle) ] \langle T \varPsi(\scriptstyle
1^{\prime} \displaystyle\sigma_{1}^{\prime})
\varPsi^{\dag}(\scriptstyle 2 \displaystyle\sigma_{2})e^{-V}
\rangle
\nonumber\\
=&-\delta_{12} [\hat{a}_{1}(\sigma_{1}\sigma_{2})\Phi]-
\sigma_{1}\delta_{11^{\prime}}v_{1}^{02} \langle T
\varPsi^{\dag}(\scriptstyle 1^{\prime}
\displaystyle\bar{\sigma}_{1})\tau^{x} \varPsi^{\dag}(\scriptstyle
2 \displaystyle\sigma_{2})e^{-V} \rangle
\label{eq:3.11}\\
&-\sigma_{1}
\left(
\frac{\delta\Phi}{\delta v_{1}^{02}}+
\frac{\delta}{\delta v_{1}^{02}}
\right)
\langle T
\varPsi^{\dag}(\scriptstyle 1^{\prime} \displaystyle\bar{\sigma_{1}})
\hat{t}(\scriptstyle 1^{\prime}1 \displaystyle)
i\tau^{y}
\varPsi^{\dag}(\scriptstyle 2 \displaystyle\sigma_{2})e^{-V}
\rangle.
\nonumber
\end{align}

We see that the equation for the GF $\langle T\varPsi\varPsi^{\dag}
e^{-V}\rangle$ contains the anomalous GF $\langle T\varPsi^{\dag}
\varPsi^{\dag} e^{-V}\rangle$. Then, it is necessary to write the
equation for it, too.

Let us introduce the matrix GF:
\begin{equation}
\mathcal{L}
(\scriptstyle\underline{1}\,\underline{2}\,\displaystyle)=
-\begin{pmatrix}
\langle T
\varPsi(\scriptstyle 1 \displaystyle\sigma_{1})
\varPsi^{\dag}(\scriptstyle 2 \displaystyle\sigma_{2})e^{-V}
\rangle
&\langle T
\varPsi(\scriptstyle 1 \displaystyle\sigma_{1})
\varPsi(\scriptstyle 2 \displaystyle\sigma_{2})e^{-V}
\rangle \\
\langle T
\varPsi^{\dag}(\scriptstyle 1 \displaystyle\sigma_{1})
\varPsi^{\dag}(\scriptstyle 2 \displaystyle\sigma_{2})e^{-V}
\rangle
&\langle T
\varPsi^{\dag}(\scriptstyle 1 \displaystyle\sigma_{1})
\varPsi(\scriptstyle 2 \displaystyle\sigma_{2})e^{-V}
\rangle
\end{pmatrix}.
\label{eq:3.12}
\end{equation}
The underlined numerical index $\underline{\scriptstyle
1\displaystyle}\,$ in the left part represents the cumulative
index, containing the space-time point $\scriptstyle
1\displaystyle\,$, the spin $\sigma_{1}$, the spinor index
$\alpha_{1}$ and one more index $\nu_{1}$, accepting two values,
specifying the matrix elements (\ref{eq:3.12}), so that
\begin{equation}
\scriptstyle\underline{1}\displaystyle\,
=\{\scriptstyle 1\displaystyle
\sigma_{1}\alpha_{1}\nu_{1}\}.
\label{eq:3.13}
\end{equation}
The matrix $\mathcal{L}
(\scriptstyle\underline{1}\,\underline{2}\,\displaystyle)$ is an
$8\times8$ matrix with respect to the collection of the discrete
indexes. A matrix of such a rank appears automatically in the
Hubbard model. Its arising is described "normal" states
(without the Cooper's pairs) but also with broken symmetries as well.

The set of four equations for the GFs in (\ref{eq:3.12}) could be
written as a single matrix equation:
\begin{equation}
\left[
L_{0V}^{-1}
(\scriptstyle\underline{1}\,\underline{1}\,^{\prime}\,\displaystyle)-
(\hat{A}\Phi Y)
(\scriptstyle\underline{1}\,\underline{1}\,^{\prime}\,\displaystyle)-
(\hat{A}Y)
(\scriptstyle\underline{1}\,\underline{1}\,^{\prime}\,\displaystyle)
\right]
\mathcal{L}
(\scriptstyle\underline{1}\,^{\prime}\underline{2}\,\,\displaystyle)=
(\hat{A}\Phi)
(\scriptstyle\underline{1}\,\underline{2}\,\displaystyle).
\label{eq:3.14}
\end{equation}
Here we introduced the operator matrix
\begin{equation}
\hat{A}
(\scriptstyle\underline{1}\,\underline{2}\,\displaystyle)=
\delta_{12}
\begin{pmatrix}
\hat{a}_{1}(\sigma_{1}\sigma_{2}) &
-\sigma_{1}\delta_{\bar{\sigma_{1}}\sigma_{2}}
i\tau^{y}
\frac{\displaystyle\delta}{\displaystyle\delta v_{1}^{02}} \\
-\sigma_{1}\delta_{\bar{\sigma_{1}}\sigma_{2}}
i\tau^{y}
\frac{\displaystyle\delta}{\displaystyle\delta v_{1}^{20}} &
\hat{a}_{1}(\sigma_{2}\sigma_{1})
\end{pmatrix},
\label{eq:3.15}
\end{equation}
where each element represents the $2\times2$ matrix with respect
to the spinor indexes, hidden in the Pauli matrices and the matrix
$\hat{a}_{1}$, having the variational derivatives with respect to
the fluctuating fields as its elements. Besides, the equation
(\ref{eq:3.14}) contains the matrix
\begin{equation}
Y(\scriptstyle\underline{1}\,\underline{2}\,\displaystyle)=
\begin{pmatrix}
\hat{t}(\scriptstyle 12 \displaystyle) & 0 \\
0 & -\hat{\tilde{t}}(\scriptstyle 12 \displaystyle)
\end{pmatrix}.
\label{eq:3.16}
\end{equation}
The value $L_{0V}^{-1}$ represents the double-row matrix
\begin{equation}
L_{0V}^{-1}
(\scriptstyle\underline{1}\,\underline{2}\displaystyle\,)=
\begin{pmatrix}
G_{0V}^{-1}
(\scriptstyle 1\displaystyle\sigma_{1},\,
\scriptstyle 2\displaystyle\sigma_{2}) &
\sigma_{1}\delta_{\bar{\sigma_{1}}
\sigma_{2}}\delta_{12}\tau^{x}v_{1}^{02} \\
-\sigma_{1}\delta_{\bar{\sigma_{1}}\sigma_{2}}
\delta_{12}\tau^{x}v_{1}^{20} &
\widetilde{G}_{0V}^{-1}
(\scriptstyle 1\displaystyle\sigma_{1},\,
\scriptstyle 2\displaystyle\sigma_{2})
\end{pmatrix},
\label{eq:3.17}
\end{equation}
where $G_{0V}^{-1}$ is given by the expression (\ref{eq:3.6}), and $%
\widetilde{G}_{0V}^{-1}$ by its transposition:
\[
\widetilde{G}_{0V}^{-1}
(\scriptstyle 1 \displaystyle\sigma_{1},\,
\scriptstyle 2 \displaystyle\sigma_{2})=
\left\{
\left(
-\frac{\partial}{\partial\tau_{1}}+E_{1}^{\sigma_{1}}
\right)
\delta_{\sigma_{1}\sigma_{2}}+
W_{1}^{\sigma_{1}}\delta_{\sigma_{1}\sigma_{2}}+
v_{1}^{\sigma_{1}\bar{\sigma_{1}}}\tau^{0}
\delta_{\bar{\sigma_{1}} \sigma_{2}}
\right\}
\delta_{12}.
\]

If replace back in equation (\ref{eq:3.14}) term with
by the mixture GFs one can see that the matrix equation (\ref{eq:3.14})
is equivalent that derived by Plakida \cite{plak1,plak2}.
The equation (\ref{eq:3.14}) is of the same type of the equation
for the single particle GF, that we derived for the Hubbard model in the limit  $%
U=\infty$ \cite{iz1} and for the Heisenberg model as well. In the above
models the matrix $\hat{A}$ degenerated into a scalar, but now it
is a matrix with respect to the discrete indexes $\alpha$ and
$\nu$, likewise the other values in (\ref{eq:3.15}). By virtue of
the noted similarity of the equation (\ref{eq:3.14}) with the
respective equations of the models considered before we could
expect the same structure in the solutions of these equations, in
particular the multiplicative character of the electronic GFs. Let
us represent them as a product of the propagator $L$ and the
terminal $\Pi$ parts, respectively, namely:
\begin{equation}
\mathcal{L}
(\scriptstyle\underline{1}\,\underline{2}\,\displaystyle)=
L(\scriptstyle\underline{1}\,\underline{1}\,^{\prime}\displaystyle)
\Pi(\scriptstyle\underline{1}\,^{\prime}\underline{2}\,\displaystyle).
\label{eq:3.18}
\end{equation}
The propagator part satisfies the Dyson equation
\begin{equation}
L^{-1}
(\scriptstyle\underline{1}\,\underline{2}\,\displaystyle)=
L_{0V}^{-1}
(\scriptstyle\underline{1}\,\underline{2}\,\displaystyle)-
\Sigma
(\scriptstyle\underline{1}\,\underline{2}\,\displaystyle).
\label{eq:3.19}
\end{equation}
Let us represent the equation for the self-energy part like the sum of the
two terms:
\begin{equation}
\Sigma
(\scriptstyle\underline{1}\,\underline{2}\,\displaystyle)=
\Sigma^{\prime}
(\scriptstyle\underline{1}\,\underline{2}\,\displaystyle)+
(\Pi Y)
(\scriptstyle\underline{1}\,\underline{2}\,\displaystyle),
\label{eq:3.20}
\end{equation}
which took place for the models considered before. Then, inserting (\ref%
{eq:3.19}) and (\ref{eq:3.20}) in (\ref{eq:3.18}) and comparing
with the initial equation (\ref{eq:3.14}), we can obtain the two
equations for $\Pi $ and $\Sigma^{\prime}$:
\begin{equation}
\Pi
(\scriptstyle\underline{1}\,\underline{2}\,\displaystyle)=
(\hat{A}\Phi)
(\scriptstyle\underline{1}\,\underline{2}\,\displaystyle)+
(YL)
(\scriptstyle\underline{4}\,^{\prime}\underline{3}\,^{\prime}\displaystyle)
\hat{A}
(\scriptstyle\underline{1}\,\underline{4}\,^{\prime}\displaystyle)
\Pi
(\scriptstyle\underline{3}\,^{\prime}\underline{2}\displaystyle),
\label{eq:3.21}
\end{equation}
\begin{equation}
\Sigma^{\prime}
(\scriptstyle\underline{1}\,\underline{2}\,\displaystyle)=
-(YL)
(\scriptstyle\underline{4}\,^{\prime}\underline{3}\,^{\prime}\displaystyle)
\hat{A}
(\scriptstyle\underline{1}\,\underline{4}\,^{\prime}\displaystyle)
\left(
L_{0V}^{-1}
(\scriptstyle\underline{3}\,^{\prime}\underline{2}\,\displaystyle)-
\Sigma^{\prime}
(\scriptstyle\underline{3}\,^{\prime}\underline{2}\,\displaystyle)
\right).
\label{eq:3.22}
\end{equation}
In obtaining these equation we have taken into account the
identity
\begin{equation}
(\hat{A}L)
(\scriptstyle\underline{1}\,\underline{2}\,\displaystyle)=
-L(\scriptstyle\underline{1}\,^{\prime}\underline{2}\,^{\prime}\displaystyle)
\left[
\hat{A}
(\scriptstyle\underline{1}\,\underline{1}\,^{\prime}\displaystyle)
L^{-1}
(\scriptstyle\underline{2}\,^{\prime}\underline{3}\,^{\prime}\displaystyle)
\right]
L(\scriptstyle\underline{3}\,^{\prime}\underline{2}\,\displaystyle),
\label{eq:3.23}
\end{equation}
which is the generalization of the well known identity expressing
the differentiation of a GF through the differentiation of its
inverse:
\[
\frac{\delta G}{\delta v}=-G \frac{\delta G^{-1}}{\delta v} G.
\]

The equations (\ref{eq:3.21}) and (\ref{eq:3.22}) for the terminal
and self-energy parts of the GF have a structure analogous to the respective
equations of the other models. These are the
equations for the variational
derivatives for $\Pi$ and $\Sigma^{\prime}$. The contribution $%
\Sigma^{\prime}$ in the self-energy part $\Sigma$ is not cutable through the
``line of the interaction", representing the value $Y$. The cutable part $%
\Sigma$ has been already extracted in the equation (\ref{eq:3.20}) like the
second contribution.

From the set of the equations (\ref{eq:3.18}) -- (\ref{eq:3.20})
it follows an important consequence, which could be represented in
the form of the following equation for the GF $\mathcal{L}$:
\begin{equation}
\mathcal{L}=\mathcal{L}^{\prime }+
\mathcal{L}^{\prime }Y\mathcal{L}.
\label{eq:3.24}
\end{equation}
Here $\mathcal{L}^{\prime }$ is determined by the two relations:
\[
\mathcal{L}^{\prime }=L^{\prime }\Pi ,\qquad
L^{\prime \ -1}=L_{0V}^{-1}-\Sigma ^{\prime }\,.
\]

The solution of the equation (\ref{eq:3.24}) could be written as:
\begin{equation}
\mathcal{L}
(\scriptstyle\underline{1}\,\underline{2}\,\displaystyle)=
\left[
\mathcal{L}^{\prime  -1}-Y
\right] ^{-1}
(\scriptstyle\underline{1}\,\underline{2}\,\displaystyle),
\label{eq:3.25}
\end{equation}
where
\begin{equation}
\mathcal{L}^{\ \prime \ -1}=
\Pi ^{-1}
\left(
L^{-1}_{0V}-\Sigma ^{\prime}
\right).
\label{eq:3.26}
\end{equation}
As it follows from the definition, the value $\mathcal{L}^{\prime
}$ is not cutable through the line $Y$, therefore the equation
(\ref{eq:3.24}) for the GF is the Larkin's equation, expressing a
GF through an irreducible part (with respect to a line of
``interaction"). From this equation it follows the locator
representation (\ref{eq:3.25}) for the electronic GF, also.

So, this issue is a diagrammatic justification of the multiplicative 
representation (\ref{eq:3.18}) for one-particle electron GF.
Similar representations for one-particle GFs in other models of
strongly correlated electron and spin systems was discussed in details
in a review \cite{iz4}.
The equations (\ref{eq:3.21}) and (\ref{eq:3.22}) could be solved by
iterations. At the first orders with respect to $Y$ we obtain:
\begin{equation}
\Pi
(\scriptstyle\underline{1}\,\underline{2}\,\displaystyle)=
\hat{A}
(\scriptstyle\underline{1}\,\underline{2}\,\displaystyle)
\Phi+(YL)
(\scriptstyle\underline{4}\,^{\prime }\underline{3}\,^{\prime }\displaystyle)
\hat{A}
(\scriptstyle\underline{1}\,\underline{4}\,^{\prime }\displaystyle)
\hat{A}
(\scriptstyle\underline{3}\,^{\prime }\underline{2}\,\displaystyle)
\Phi +...\,,
\label{eq:3.27}
\end{equation}
\begin{equation}
\Sigma ^{\prime }
(\scriptstyle\underline{1}\,\underline{2}\,\displaystyle)= -(YL)
(\scriptstyle\underline{4}\,^{\prime}\underline{3}\,^{\prime}\displaystyle)
\left[ \hat{A}
(\scriptstyle\underline{1}\,\underline{4}\,^{\prime}\displaystyle)
L_{0V}^{-1}
(\scriptstyle\underline{3}\,^{\prime}\underline{2}\,\displaystyle)
\right] \label{eq:3.28}
\end{equation}
\[
+(YL)
(\scriptstyle\underline{4}\,^{\prime}\underline{3}\,^{\prime}\displaystyle)
(YL)
(\scriptstyle\underline{6}\,^{\prime}\underline{1}\,^{\prime}\displaystyle)
\left[
\hat{A}
(\scriptstyle\underline{1}\,\underline{4}\,^{\prime}\displaystyle)
L_{0V}^{-1}
(\scriptstyle\underline{1}\,^{\prime}\underline{2}\,^{\prime}\displaystyle)
\right]
L(\scriptstyle\underline{2}\,^{\prime}\underline{5}\,^{\prime}\displaystyle)
\left[
\hat{A}
(\scriptstyle\underline{3}\,^{\prime}\underline{6}\,^{\prime}\displaystyle)
L_{0V}^{-1}
(\scriptstyle\underline{5}\,^{\prime }\underline{2}\,\displaystyle)
\right]+...\,,
\nonumber
\]

In (\ref{eq:3.27}) the operator $\hat{A}$, acting on $\Phi$,
brings the mean value of the diagonal and $b$-operators; a
repeated action of the operator $\hat{A}$ will produce bosonic GFs
of the different types. An action of the operator on
$L_{\upsilon}^{-1}$ will result in expressions composed of
different $\delta$-symbols. The problem is contained in the
multiplication of the matrices in the equations (\ref{eq:3.27})
and (\ref{eq:3.28}), taking into account that the matrix
$\hat{A}(\scriptstyle
\underline{1}\,\underline{2}\,\displaystyle)$ contains
derivatives, which should act on the corresponding values. To
fulfil the matrix multiplication accounting for the operator
character of the several factors, we rewrite the expressions
(\ref{eq:3.27}) and (\ref{eq:3.28}) in another form:
\begin{equation}
\Pi
(\scriptstyle\underline{1}\,\underline{2}\,\displaystyle)=
\hat{A}
(\scriptstyle\underline{1}\,\underline{2}\,\displaystyle)
\Phi+\hat{A}
(\scriptstyle\underline{1}\,\underline{4}\,^{\prime}\displaystyle)
(YL)
(\scriptstyle\underline{4}\,^{\prime}\underline{3}\,^{\prime}\displaystyle)
\hat{A}
(\scriptstyle\underline{3}\,^{\prime}\underline{2}\,\displaystyle)
\Phi+...\,,
\label{eq:3.29}
\end{equation}
\begin{equation}
\Sigma^{\prime}
(\scriptstyle\underline{1}\,\underline{2}\,\displaystyle)=
-\hat{A}
(\scriptstyle\underline{1}\,\underline{4}\,^{\prime}\displaystyle)
(YL)
(\scriptstyle\underline{4}\,^{\prime}\underline{3}\,^{\prime}\displaystyle)
L_{V}^{-1} (\scriptstyle
\underline{3}\,^{\prime}\underline{2}\,\displaystyle)
\label{eq:3.30}
\end{equation}
\[
+\hat{A}
(\scriptstyle \underline{1}\,\underline{4}\,^{\prime}\displaystyle)
(YL)
(\scriptstyle \underline{4}\,^{\prime}\underline{3}\,^{\prime}\displaystyle)
\hat{A}
(\scriptstyle \underline{3}\,^{\prime}\underline{6}\,^{\prime}\displaystyle)
(YL)
(\scriptstyle \underline{6}\,^{\prime}\underline{1}\,^{\prime}\displaystyle)
L_{V}^{-1}
(\scriptstyle \underline{1}\,^{\prime}\underline{2}\,^{\prime}\displaystyle)
L(\scriptstyle \underline{2}\,^{\prime}\underline{5}\,^{\prime}\displaystyle)
L_{V}^{-1}
(\scriptstyle \underline{5}\,^{\prime}\underline{2}\,\displaystyle)
+...\,,
\]
In these expressions all the factors are arranged in the order of
the matrix multiplication, but we should not forget which factors
the derivatives of the matrix $\hat{A}$ act on.


\section{Iteration equation for the self-energy and terminal part}
According to definition (\ref{eq:3.12}), the electronic GF
$\mathcal{L}$ takes into account the possibility of states with
coupled electrons. In this paper we shall consider the normal
system, described completely by the matrix element of the
electronic GF $\mathcal{L}(\scriptstyle 12 \displaystyle)$, namely
\begin{equation}
\mathcal{G}
(\scriptstyle 1 \displaystyle\sigma_{1}\,
\scriptstyle 2 \displaystyle\sigma_{2}\,)
\equiv
\mathcal{L}^{11}
(\scriptstyle 1 \displaystyle\sigma_{1}\,
\scriptstyle 2 \displaystyle\sigma_{2}\,).
\label{eq:4.1}
\end{equation}

The normal GF $\mathcal{G}$ can be looked in the standard
multiplicative form
\begin{equation}
\mathcal{G}=G\Lambda.
\label{eq:4.2}
\end{equation}
with $G$ obeying the Dyson equation
\begin{equation}
G^{-1}=G_{0V}^{-1}-\Sigma ,
\label{eq:4.3}
\end{equation}
and the self-energy part being a sum of two terms, uncutable
$\Sigma^{\prime}$ and cutable $\Lambda\hat{t}$:
\begin{equation}
\Sigma=\Sigma^{\prime}+\Lambda\hat{t}.
\label{eq:4.4}
\end{equation}
In equations (\ref{eq:4.2}) -- (\ref{eq:4.4}) all quantities are
$2\times2$ matrices with respect to spinor indexes, with arguments
of the type $\mathcal{G} (\scriptstyle 1 \displaystyle\sigma_{1}\,
\scriptstyle 2 \displaystyle\sigma_{2}\,)$.

Iterations in general equations (\ref{eq:3.21}) and
(\ref{eq:3.22}) allow to get series for $\Sigma^{\prime}$ and
$\Lambda$, determined by equations (\ref{eq:4.2}) --
(\ref{eq:4.4}). Calculations of these series are done in Appendix
B, and here we present the results within the limit of the first
two orders. We have, for the zeroth order of $\Lambda$:
\begin{equation}
\Lambda_{0}^{\sigma}(k)=
\begin{pmatrix}
1-\langle n^{\bar{\sigma}} \rangle & 0 \\
0 & \langle n^{\bar{\sigma}} \rangle
\end{pmatrix},
\label{eq:4.5}
\end{equation}
where
\begin{equation}
\langle n^{\sigma} \rangle=
\langle X_{i}^{\sigma\sigma}
+ X_{i}^{22} \rangle=
\langle c_{i\sigma}^{\dag}c_{i\sigma} \rangle
\label{eq:4.6}
\end{equation}
is the average number of electrons on a site with spin $\sigma$.
The first order correction for $\Lambda$ is the following
\begin{equation}
\Lambda_{1}^{\sigma}(k)=
\begin{pmatrix}
\lambda_{1}^{\sigma}(k) & \lambda_{2}^{\sigma}(k) \\
-\lambda_{1}^{\sigma}(k) & -\lambda_{2}^{\sigma}(k)
\end{pmatrix},
\label{eq:4.7}
\end{equation}
where
\begin{align}
\lambda_{1}^{\sigma}(k)=& -\sum\limits_{q}
\varepsilon(\bm{k}+\bm{q})\biggl[
(G_{11}^{\sigma}+G_{21}^{\sigma})(k+q)
\mathcal{N}^{\bar{\sigma}\bar{\sigma}}(q)
\label{eq:4.8}\\
&+(G_{11}^{\bar{\sigma}}+G_{21}^{\bar{\sigma}})(k+q)
\mathcal{D}^{\sigma\bar{\sigma}}(q)+
(G_{21}^{\bar{\sigma}}+G_{22}^{\bar{\sigma}})(-k-q)
\mathcal{D}^{02}(q)\biggr],
\nonumber
\end{align}
\begin{align}
\lambda_{2}^{\sigma}(k)&= \sum\limits_{q}
\varepsilon(\bm{k}+\bm{q})\biggl[
(G_{12}^{\sigma}+G_{22}^{\sigma})(k+q)
\mathcal{N}^{\bar{\sigma}\bar{\sigma}}(q)
\label{eq:4.9}\\
&+(G_{12}^{\bar{\sigma}}+G_{22}^{\bar{\sigma}})(k+q)
\mathcal{D}^{\sigma\bar{\sigma}}(q)+
(G_{11}^{\bar{\sigma}}+G_{12}^{\bar{\sigma}})(-k-q)
\mathcal{D}^{02}(q)\biggr].
\nonumber
\end{align}
The quantities $\mathcal{N}^{\bar{\sigma}\bar{\sigma}}(k)$,
$\mathcal{D}^{\sigma\bar{\sigma}}(k)$ and $\mathcal{D}^{02}(k)$
are the Fourier transforms of the bosonic GFs, determined by
relations (\ref{eq:2.20}) -- (\ref{eq:2.22}) with 4-momentum $q$.
Here $\varepsilon(\bm{k})$ is the Fourier transform of
$t_{i_{1}i_{2}}$, which is actually the bare electron energy in
the lattice.

The contribution of the first order in $\Sigma^{\prime}$ is given
by:
\begin{equation}
\Sigma_{1}^{\prime \sigma}=
-\eta^{\sigma}
\begin{pmatrix}
1 & -1 \\
-1 & 1
\end{pmatrix},
\label{eq:4.10}
\end{equation}
where
\begin{equation}
\eta^{\sigma}=
\sum\limits_{k}\varepsilon(\bm{k})
\left[
G_{11}^{\bar{\sigma}}(k)-G_{22}^{\bar{\sigma}}(k)
\right].
\label{eq:4.11}
\end{equation}
The second order correction is equal to:
\begin{equation}
\Sigma_{2}^{\prime\sigma}=
\begin{pmatrix}
\varphi_{1}^{\sigma}(k) &
\varphi_{2}^{\sigma}(k) \\
-\varphi_{1}^{\sigma}(k) &
-\varphi_{2}^{\sigma}(k)
\end{pmatrix},
\label{eq:4.12}
\end{equation}
where
\begin{align}
\varphi_{1}^{\sigma}(k)&= \sum\limits_{q}\sum\limits_{k_{1}}
\varepsilon(\bm{k+q})\varepsilon(\bm{k_{1}+q}) \biggl[
G_{11}^{\sigma}(k_{1}) g^{\bar{\sigma}}(k+q)
G_{11}^{\bar{\sigma}}(k_{1}+q)
\label{eq:4.13}\\
&+G_{11}^{\bar{\sigma}}(k_{1})
g^{\sigma}(k+q)
G_{11}^{\bar{\sigma}}(k_{1}+q)+
\sum\limits_{\sigma^{\prime}}
G_{22}^{\bar{\sigma}}(-k_{1})
g^{\sigma^{\prime}}(k+q)
G_{11}^{\bar{\sigma}^{\prime}}(-k_{1}-q)
\biggr],
\nonumber
\end{align}
and the quantity $\varphi_{2}^{\sigma}(k)$ is given by a change of
spinor indexes $1\leftrightarrow2$ in (\ref{eq:4.13}). Here
$g^{\sigma}(k)$ is a linear combination of the matrix elements of
the electronic GF:
\begin{equation}
g^{\sigma}(k)=
G_{11}^{\sigma}(k)+G_{21}^{\sigma}(k)-
G_{12}^{\sigma}(k)-G_{22}^{\sigma}(k)
\label{eq:4.14}
\end{equation}

Finally we write down the second order contribution in the cutable
part of $\Sigma$, that is, the  expression for
$\Sigma_{red}^{\sigma}\equiv\Lambda^{\sigma}\hat{t}^{\sigma}$.
Because in momentum representation $\hat{t}^{\sigma}$ is equal to
$\varepsilon(\bm{k})\Im$, with $\Im$ being the $2\times2$ matrix
determined in (\ref{eq:2.8}), we find, according to
(\ref{eq:4.8}):
\begin{equation}
\Sigma_{red}^{\sigma}=
\lambda^{\sigma}(k)
\varepsilon(\bm{k})
\begin{pmatrix}
1 & 1 \\
-1 & -1
\end{pmatrix},
\label{eq:4.15}
\end{equation}
where
\begin{align}
&\lambda^{\sigma}(k)= \lambda_{1}^{\sigma}(k)+
\lambda_{2}^{\sigma}(k)
\nonumber\\
=-\sum\limits_{q} \varepsilon(\bm{k+q})&\biggl[
g^{\sigma}(k+q)\mathcal{N}^{\bar{\sigma}\bar{\sigma}}(q)+
g^{\bar{\sigma}}(k+q)\mathcal{D}^{\sigma\bar{\sigma}}(q)
\label{eq:4.16}\\
&+\widetilde{g}^{\bar{\sigma}}(k+q)\mathcal{D}^{02}(q)
\biggr],
\nonumber
\end{align}
\begin{equation}
\widetilde{g}^{\sigma}(k)=
G_{21}^{\sigma}(-k)+G_{22}^{\sigma}(-k)-
G_{11}^{\sigma}(-k)-G_{12}^{\sigma}(-k),
\label{eq:4.17}
\end{equation}
and
\[
\widetilde{g}^{\sigma}(k)=-g^{\sigma}(-k).
\]
We see that the correction $\Sigma_{1}^{\prime\sigma}$ depends
neither on momentum nor on frequency and determines only a shift
of electron spectrum, but it depends on spin. The second order
corrections $\Sigma_{2}^{\prime\sigma}(k)$ and
$\Sigma_{red}^{\sigma}(k)$ depend both on momentum and frequency.
The contribution $\Sigma_{red}^{\sigma}$ is determined by the
interaction of electrons with bosonic excitations, while
$\Sigma_{2}^{\prime\sigma}$ is determined by electronic GFs only.


\section{Mean field approximation}
The simplest approximation of a mean field type is the Hubbard-I,
which takes into account a term in $\Sigma$ equal to
$\Lambda_{0}\hat{t}$. To it, one can add a first order term
$\Sigma_{1}^{\prime}$, not depending on frequency. The second
order correction $\Sigma_{2}^{\prime}$ depends on the frequency,
however we shall try to extract from it a static part by the
following ansatz.

Let us consider that in $\Sigma_{2}^{\prime}$ both expressions for
$\varphi_{1}^{\sigma}(k)$, $\varphi_{2}^{\sigma}(k)$ include a factor
$\varepsilon(\bm{k+q})$ in the summation over $\bm{q}$. So it can
be factorized in the nearest neighbor approximation, and a term
proportional to $\varepsilon(\bm{k})$ can be taken out from the
static part of $\Sigma_{2}^{\prime}$ for the cubic lattice. Thus
in the static approximation $\Sigma_{2}^{\prime}$ can be
approximated by the expression
\begin{equation}
\Sigma_{2}^{\prime}(\bm{k})=
\begin{pmatrix}
p_{1}^{\sigma} & p_{2}^{\sigma} \\
-p_{1}^{\sigma}&-p_{2}^{\sigma}
\end{pmatrix}
\varepsilon(\bm{k}).
\label{eq:5.1}
\end{equation}
Here $p_{1}^{\sigma}$ and $p_{2}^{\sigma}$ are some spin dependent
constants. Their expressions can be explicitly written out, but we
will not do it, because we shall try to calculate them from some
general conditions for electronic GFs, which should be satisfied.
Such conditions were formulated in works by Mancini and coworkers
(see general discussion in paper \cite{man2} and refs. therein), where it
is developed a method using linearized equation of motion for
composite operators. The condition is demanding that the
electronic GF $\mathcal{G}_{12}$ is equal to zero when arguments
coincide. Below we will use this idea for the determination of
unknown parameters $p_{1}^{\sigma}$ and $p_{2}^{\sigma}$.

First we write down the self-energy part in an approximation which
includes the Hubbard-I term, the first order correction
$\Sigma^{\prime}$ (\ref{eq:4.10}) and $\Sigma_{2}^{\prime}$ in the
form (\ref{eq:5.1}). All these three contributions give
$\Sigma_{MF}$, corresponding to a mean field approximation. So we
have:
\begin{equation}
\Sigma_{MF}^{\sigma}(\bm{k})=
-\eta^{\sigma}
\begin{pmatrix}
1 & -1 \\
-1 & 1
\end{pmatrix}+
\begin{pmatrix}
1-\langle n^{\bar{\sigma}}\rangle+p_{1}^{\sigma}&
1-\langle n^{\bar{\sigma}}\rangle+p_{2}^{\sigma}
\\
\langle n^{\bar{\sigma}}\rangle-p_{1}^{\sigma}&
\langle n^{\bar{\sigma}}\rangle-p_{2}^{\sigma}
\end{pmatrix}
\varepsilon(\bm{k})\,.
\label{eq:5.2}
\end{equation}
It is clear that the first term is responsible for a shift of the
Hubbard subbands, and the second one for a renormalization of
their widths. The propagator part of the GF in the mean field
approximation is determined by a matrix equation:
\[
\left[G^{\sigma}(k)\right]^{-1}=
\left[G_{0}^{\sigma}(k)\right]^{-1}-
\Sigma_{MF}^{\sigma}(k).
\]
We look for a solution of the form
\begin{equation}
G_{\alpha\beta}^{\sigma}(k)=
\frac{\displaystyle
(\mathcal{A}_{\scriptscriptstyle 1 \displaystyle}^{\sigma})_{\alpha\beta}
(\bm{k})}
{\displaystyle i\omega_{n}-
E_{\scriptscriptstyle 1 \displaystyle}^{\sigma}(\bm{k})}+
\frac{\displaystyle
(\mathcal{A}_{\scriptscriptstyle 2 \displaystyle}^{\sigma})_{\alpha\beta}
(\bm{k})}
{\displaystyle i\omega_{n}-
E_{\scriptscriptstyle 2 \displaystyle}^{\sigma}(\bm{k})}.
\label{eq:5.3}
\end{equation}
The poles $E_{\scriptscriptstyle m
\displaystyle}^{\sigma}(\bm{k})$ and their residues
$(\mathcal{A}_{\scriptscriptstyle m
\displaystyle}^{\sigma})_{\alpha\beta} (\bm{k})$ are written in
the form:
\begin{equation}
\left.
\begin{split}
&(\mathcal{A}_{\scriptscriptstyle 1,2 \displaystyle}
^{\sigma})_{11}(\bm{k})=
\frac{\displaystyle 1}{\displaystyle 2}
\left[
1 \pm \frac{\displaystyle r^{\sigma}(\bm{k})}
{\displaystyle 2Q^{\sigma}(\bm{k})}
\right]
\\
&(\mathcal{A}_{\scriptscriptstyle 1,2 \displaystyle}
^{\sigma})_{22}(\bm{k})=
\frac{\displaystyle 1}{\displaystyle 2}
\left[
1 \mp \frac{\displaystyle r^{\sigma}(\bm{k})}
{\displaystyle 2Q^{\sigma}(\bm{k})}
\right]
\\
&(\mathcal{A}_{\scriptscriptstyle 1,2 \displaystyle}
^{\sigma})_{12}(\bm{k})=
\mp \frac{\displaystyle \eta^{\sigma}
+(1-\langle n^{\bar{\sigma}}\rangle+p_{2}^{\sigma})
\varepsilon(\bm{k})}
{2Q^{\sigma}(\bm{k})}
\\
&(\mathcal{A}_{\scriptscriptstyle 1,2 \displaystyle}
^{\sigma})_{21}(\bm{k})=
\mp \frac{\displaystyle \eta^{\sigma}
+(\langle n^{\bar{\sigma}}\rangle-p_{1}^{\sigma})
\varepsilon(\bm{k})}
{2Q^{\sigma}(\bm{k})}
\end{split}
\right\},
\label{eq:5.4}
\end{equation}
\begin{equation}
E_{\scriptscriptstyle 1,2 \displaystyle}^{\sigma}(\bm{k})=
R^{\sigma}(\bm{k})\mp Q^{\sigma}(\bm{k}).
\label{eq:5.5}
\end{equation}
Here
\[
r^{\sigma}(\bm{k})=
U-[1-2\langle n^{\bar{\sigma}}\rangle+p_{1}^{\sigma}+p_{2}^{\sigma}]
\varepsilon(\bm{k}),
\]
whilst expressions for $R^{\sigma}(\bm{k})$ and
$Q^{\sigma}(\bm{k})$ will be written later.

The electronic GF $\mathcal{G}^{\sigma}$ in the mean field
approximation is found with the help of the general relation
(\ref{eq:4.2})
\[
\mathcal{G}^{\sigma}(k)=G^{\sigma}(k)\Lambda_{0}^{\sigma}(k),
\]
where $\Lambda_{0}^{\sigma}(k)$ is given by the matrix
(\ref{eq:4.5}).

The electronic GF depends on parameters $\mu$, $\eta^{\sigma}$,
$\langle n^{\sigma}\rangle$, $p_{1}^{\sigma}$ and
$p_{2}^{\sigma}$, which must be determined in a self consistent
way from the equations
\[
\sum\limits_{\sigma}
\langle n_{i}^{\sigma}\rangle=n_{i},\qquad
\langle n_{i}^{\sigma}\rangle=
\sum\limits_{\alpha\beta}
\mathcal{G}_{\alpha\beta}^{\sigma}
(i,\tau\,;\,i,\tau+0),
\]
and also from equation (\ref{eq:4.11}), determining the parameter
$\eta^{\sigma}$. Parameters $p_{1}^{\sigma}$ and $p_{2}^{\sigma}$
will be determined [10] from conditions which follow from the
properties of $X$ operators, namely:
\begin{equation}
\left.
\begin{split}
&\mathcal{G}_{12}^{\sigma}
(i,\tau\,;\,i,\tau+0)=
\langle
\bar{\sigma}X_{i}^{2\bar{\sigma}}X_{i}^{0\sigma}
\rangle=0
\\
&\mathcal{G}_{21}^{\sigma}
(i,\tau\,;\,i,\tau+0)=
\langle
X_{i}^{\sigma 0}X_{i}^{\bar{\sigma} 2}\bar{\sigma}
\rangle=0
\end{split}
\right\}.
\label{eq:5.6}
\end{equation}
Thus a complete system of equations for all five parameters can be
written in the form:
\begin{equation}
\langle n^{\sigma}\rangle+
\langle n^{\bar{\sigma}}\rangle=n,
\label{eq:5.7}
\end{equation}
\begin{equation}
\langle n^{\sigma}\rangle=
\sum\limits_{k}
\left[
\mathcal{G}_{11}^{\sigma}(k)+
\mathcal{G}_{22}^{\sigma}(k)
\right],
\label{eq:5.8}
\end{equation}
\begin{equation}
\eta^{\sigma}=
\sum\limits_{k}\varepsilon(\bm{k})
\left[
G_{11}^{\bar{\sigma}}(k)-G_{22}^{\bar{\sigma}}(k)
\right],
\label{eq:5.9}
\end{equation}
\begin{equation}
\sum\limits_{k}\mathcal{G}_{12}^{\sigma}(k)=0,
\label{eq:5.10}
\end{equation}
\begin{equation}
\sum\limits_{k}\mathcal{G}_{21}^{\sigma}(k)=0
\label{eq:5.11}
\end{equation}
(we assume homogeneous states, so all averages do not depend on
site index).

From the comparison of the last two equations we find a relation
between parameters $p_{1}^{\sigma}$ and $p_{2}^{\sigma}$:
\begin{equation}
p_{1}^{\sigma}+p_{2}^{\sigma}=
-(1-2\langle n^{\bar{\sigma}}\rangle).
\label{eq:5.12}
\end{equation}
Therefore, the parameter $p_{2}^{\sigma}$ can be replaced in all
the expressions above. Thus the equations (\ref{eq:5.4}) for the
residues of GF are:
\begin{equation}
\left.
\begin{split}
&(\mathcal{A}_{\scriptscriptstyle 1,2 \displaystyle}
^{\sigma})_{11}(\bm{k})=
\frac{\displaystyle 1}{\displaystyle 2}
\left[
1 \pm \frac{\displaystyle U}
{\displaystyle 2Q^{\sigma}(\bm{k})}
\right]
\\
&(\mathcal{A}_{\scriptscriptstyle 1,2 \displaystyle}
^{\sigma})_{22}(\bm{k})=
\frac{\displaystyle 1}{\displaystyle 2}
\left[
1 \mp \frac{\displaystyle U}
{\displaystyle 2Q^{\sigma}(\bm{k})}
\right]
\\
(\mathcal{A}_{\scriptscriptstyle 1,2 \displaystyle}
^{\sigma})_{12}&(\bm{k})=
(\mathcal{A}_{\scriptscriptstyle 1,2 \displaystyle}
^{\sigma})_{21}(\bm{k})=
\mp \frac{\displaystyle \eta^{\sigma}
+(\langle n^{\bar{\sigma}}\rangle-p_{1}^{\sigma})
\varepsilon(\bm{k})}
{2Q^{\sigma}(\bm{k})}
\end{split}
\right\}.
\label{eq:5.13}
\end{equation}
Expressions for $R^{\sigma}(\bm{k})$ and $Q^{\sigma}(\bm{k})$,
determining poles, are now equal to:
\begin{equation}
\left.
\begin{split}
&R^{\sigma}(\bm{k})=
-\sigma\frac{\displaystyle h}{\displaystyle 2}-
\eta^{\sigma}+(1+p_{1}^{\sigma}-
\langle n^{\bar{\sigma}}\rangle)
\varepsilon(\bm{k})+
\frac{\displaystyle U}{\displaystyle 2}-\mu\quad
\\
&Q^{\sigma}(\bm{k})=
\frac{\displaystyle 1}{\displaystyle 2}
\sqrt{U^{2}+4
\left[
\eta^{\sigma}+
(\langle n^{\bar{\sigma}}\rangle-p_{1}^{\sigma})
\varepsilon(\bm{k})
\right]^{2}}
\end{split}
\right\}.
\label{eq:5.14}
\end{equation}

After the replacement of parameter $p_{2}^{\sigma}$, the two
equations (\ref{eq:5.10}) and (\ref{eq:5.11}) reduce to only one,
which allows to find the unknown parameter $p_{1}^{\sigma}$.
Taking the summation over frequencies in all equations
(\ref{eq:5.8}) -- (\ref{eq:5.11}), we write our system in the
form:
\begin{equation}
\langle n^{\sigma}\rangle=
\frac{\displaystyle 1}{\displaystyle 2}
(1-\mathcal{K}_{0}^{\sigma})-
\frac{\displaystyle U}{\displaystyle 2}
\mathcal{F}_{0}^{\sigma}
(1-2\langle n^{\bar{\sigma}}\rangle),
\label{eq:5.15}
\end{equation}
\begin{equation}
\eta^{\sigma}=-U\mathcal{F}_{1}^{\bar{\sigma}},
\label{eq:5.16}
\end{equation}
\begin{equation}
\eta^{\sigma}\mathcal{F}_{0}^{\sigma}+
(\langle n^{\bar{\sigma}}\rangle-p_{1}^{\sigma})
\mathcal{F}_{1}^{\sigma}=0,
\label{eq:5.17}
\end{equation}
where we use the definitions of the paper [8]:
\begin{equation}
\mathcal{K}_{n}^{\sigma}=
\frac{\displaystyle 1}{\displaystyle 2N}
\sum\limits_{\bm{k}}
\varepsilon^{n}(\bm{k})
\left[
\mathrm{th}\left(
\frac{\displaystyle
E_{\scriptscriptstyle 1 \displaystyle}^{\sigma}(\bm{k})}
{\displaystyle 2T}
\right)+
\mathrm{th}\left(
\frac{\displaystyle
E_{\scriptscriptstyle 2 \displaystyle}^{\sigma}(\bm{k})}
{\displaystyle 2T}
\right)
\right],
\label{eq:5.18}
\end{equation}
\begin{equation}
\mathcal{F}_{n}^{\sigma}=
\frac{\displaystyle 1}{\displaystyle 2N}
\sum\limits_{\bm{k}}
\frac{\displaystyle \varepsilon^{n}(\bm{k})}
{\displaystyle 2Q^{\sigma}(\bm{k})}
\left[
\mathrm{th}\left(
\frac{\displaystyle
E_{\scriptscriptstyle 1 \displaystyle}^{\sigma}(\bm{k})}
{\displaystyle 2T}
\right)-
\mathrm{th}\left(
\frac{\displaystyle
E_{\scriptscriptstyle 2 \displaystyle}^{\sigma}(\bm{k})}
{\displaystyle 2T}
\right)
\right].
\label{eq:5.19}
\end{equation}
We have to add to them equations (\ref{eq:5.15}) -- (\ref{eq:5.17})
and equation  (\ref{eq:5.7}) for chemical potential.

The energy of the system can be found by averaging the Hamiltonian
(\ref{eq:2.6}) -- (\ref{eq:2.7}) over a Gibbs ensemble. It is
quite easy to express it by means of electronic GFs:
\begin{equation}
\frac{1}{N}\langle \mathcal{H}\rangle=
\sum\limits_{k\sigma}\varepsilon(\bm{k})
\sum\limits_{\alpha\beta}
\mathcal{G}_{\alpha\beta}^{\sigma}(k)+
U\langle X^{22}\rangle
\label{eq:5.20}
\end{equation}
where
\begin{equation}
\langle X^{22}\rangle\equiv D=
\frac{1}{2}\sum\limits_{k\sigma}
\mathcal{G}_{22}^{\sigma}.
\label{eq:5.21}
\end{equation}
After substituting here the expressions for the matrix elements
$\mathcal{G}_{\alpha\beta}^{\sigma}$, we find the expressions for
the energy $\langle \mathcal{H}\rangle$ and the double occupation
parameter $D$:
\begin{align}
&\frac{\displaystyle 1}{\displaystyle N}
\langle \mathcal{H}\rangle=
U\langle X^{22}\rangle+
\label{eq:5.22}
\\
+\sum\limits_{\sigma}
\biggl[
-\eta^{\sigma}\langle n^{\sigma}\rangle
-\frac{\displaystyle 1}{\displaystyle 2}
&\mathcal{K}_{1}^{\sigma}-
\frac{\displaystyle U}{\displaystyle 2}
\mathcal{F}_{1}^{\sigma}+
\eta^{\sigma}\mathcal{F}_{1}^{\sigma}+
(\langle n^{\bar{\sigma}}\rangle-p_{1}^{\sigma})
\mathcal{F}_{2}^{\sigma}
\biggr],\nonumber
\end{align}
\begin{equation}
\langle X^{22}\rangle\equiv D=
\frac{1}{4}\sum\limits_{\sigma}
\langle n^{\bar{\sigma}}\rangle
(1-\mathcal{K}_{0}^{\sigma}+U\mathcal{F}_{0}^{\sigma}).
\label{eq:5.23}
\end{equation}

In Fig. 1 the parameters $p_{1}$ and $p_{2}$ are plotted as
functions of electron concentration at different $U$. Such results
are typical for other fixed parameters of the system. For all
different $n$ and $U$ the parameter $p_{1}$ is positive and
$p_{2}$ is negative. A negative solution for parameter $p_{1}$ was
not found. The behavior of $p_{1}$ is rather similar to the COM1
solution for the parameter $p$ in works \cite{ave1}\cite{ave2} (COM1 is a name
authors \cite{ave1}\cite{ave2} gave for the solution with $p>0$). In Fig.2 the
concentration dependence of chemical potential is given for two
values of $U$. In the same figure a COM1 solution, that we found
from equations of paper \cite{ave1}, is presented for two variants of
density of states for the bare electron band: a two-dimensional
square lattice with nearest-neighbor hopping and a model density
of states of this type:
\begin{equation}
\rho(\varepsilon)=
\begin{cases}
1, & |\varepsilon|<W/2\\
0, & |\varepsilon|>W/2
\end{cases}.
\label{eq:5.24}
\end{equation}
We see that COM and our GFA give similar results. The COM1
solution for the 2D-system and for the model density of states are
quantitatively very close, and because of this we shall use
hereafter for simplicity the model density of states
(\ref{eq:5.24}).

Slightly worse is the comparison of results for $\eta$ (Fig.3),
however there is a qualitative coincidence of COM1 and GFA
calculations. The parameter of double occupation $D$ gives  again
a satisfactory coincidence of the two approaches (Fig.4). It is
useful to show the dependence of $\eta$ on $n$ in a whole electron
concentration interval at different values of $U$ (Fig.5). When
decreasing $U$ a part of the curve denoted by dash lines
approaches to the abscess line, and when $U\to 0$ one can see that
$\eta\to 0$, as it should be in the case of noninteracting
electrons.

Calculations show that when $U$ decreases, the value of the jump
of the chemical potential at $n=1$ decreases too, and at the same
critical value $U_{c}\approx 1.73\,W$ becomes equal to zero. This
corresponds to the closing up the two Hubbard subbands, and the
insulator-metal phase transition occurs. The evolution of the
density of states of the quasiparticle spectrum when $U$ changes
is shown for two different bare density of states: the model one
(\ref{eq:5.24}), Fig.6, and the semielliptic one
\begin{equation}
\rho(\varepsilon)=
\begin{cases}
\frac{\displaystyle 4}{\displaystyle \pi W}
\sqrt{1-
\left(
\frac{\displaystyle 2\varepsilon}
{\displaystyle W}
\right)^{2}}&,\quad|\varepsilon|<W/2\\
0&,\quad|\varepsilon|>W/2
\end{cases},
\label{eq:5.25}
\end{equation}
Fig.7. In COM1 the critical value is $U_{c}\approx 1.68W$ \cite{ave1},
which is close to our value $U_{c}\approx 1.73\,W$, obtained for
the density of states (\ref{eq:5.24}).

At half-filling it is easy to get an expression for the gap
between the two Hubbard subbands with energies
$E_{\scriptscriptstyle 1 \displaystyle}(\bm{k})$  and
$E_{\scriptscriptstyle 2 \displaystyle}(\bm{k})$:
\begin{equation}
\Delta\,E=
-\left(
\frac{1}{2}+p_{1}
\right)W+
\sqrt{U^{2}+\left(\frac{1}{2}-p_{1}\right)^{2}W^{2}}.
\label{eq:5.26}
\end{equation}
From here follows the critical value $U_{c}$, when $\Delta E=0$.
It is equal to
\begin{equation}
U_{c}=\sqrt{2p_{1}}\,W,
\label{eq:5.27}
\end{equation}
so that when $U>U_{c}$ the system is an insulator, and when
$U<U_{c}$ a metal.

Compare now the two approaches for the Hubbard model: GFA and COM.
The mean field approximations in the framework of these approaches
are close to each other both as what regards the GFs structure and
physical properties of the model calculated with their help. In
both cases the electronic GF has a two-poles structure. The COM
approach includes only the parameter $p$, which has to be found
from the equation $\mathcal{G}_{12}=0$. In the GFA two parameters,
($p_{1}$ and $p_{2}$), appear, determined through two equations:
$\mathcal{G}_{12}=0$ and $\mathcal{G}_{21}=0$. Due to this pair of
equations one of these parameters can be eliminated, and as a
result we have only one parameter, $p_{1}$.

The physical meanings of the parameters $p$ and $p_{1}$ are close.
In the COM approach the parameter $p$ describes the static
fluctuation of charge, spin and pair. In GFA the parameter $p_{1}$
includes traces of static charge and spin fluctuations as well.
Corrections for the self-energy due to dynamical interaction of
electrons with bosons in both approaches practically coincide and
correspond to SCBA.

The equations for the determination of parameters $\mu$, $\langle
n^{\sigma}\rangle$, $\eta$, $p_{1}$ in GFA and $\mu$, $\langle
n^{\sigma}\rangle$, $\Delta$, $p$ in COM are rather similar, but
have different solutions. In COM at fixed external parameters
($n$, $U$, $W$) one has two solutions: with $p>0$ and $p<0$, while
in GFA there is only one solution with $p_{1}>0$ (the second
parameter $p_{2}$ is always negative, but it does not enter in the
electronic GF explicitly; but it only guarantees the satisfaction
of the two conditions $\mathcal{G}_{12}=0$ and
$\mathcal{G}_{21}=0$, simultaneously). A remarkable conclusion
follows from numerical calculations with different sets of
external parameters. In spite of some difference in GFA and COM,
the calculated quantities of the model are rather close to each
other, if in COM only COM1 solutions with $p>0$ are taken into
account. The two-poles GF of this approximation can
be used farther for the calculation of corrections to the
self-energy $\Sigma(\bm{k},\omega)$ from the dynamical
fluctuations \cite{ave3} and for bosonic GFs (magnon, plasmon, doublon),
describing these fluctuations \cite{ave4}.


\section{Boson Green's functions}
The complete system of 16 $X$ operators contains two Bose-like
operators $X_{1}^{\sigma\bar{\sigma}}$ and $X_{1}^{02}$ (and their
conjugates $X_{1}^{\bar{\sigma}\sigma}$ and $X_{1}^{20}$), which
determine the two Bose-like GFs (2.15) and (2.16).

They describe propagation of a spin flip (magnon) and a dyad
(doublon), representing the two types of the Bose-like collective
modes. These GFs could be represented as the variational
derivatives of $Z[V]$ with respect to the fluctuating fields:
\begin{equation}
\mathcal{D}^{\sigma\bar{\sigma}}
(\scriptstyle 12 \displaystyle)=
-\frac{\delta^{2}\Phi}
{\delta v_{1}^{\sigma\bar{\sigma}}v_{2}^{\bar{\sigma}\sigma}}\,,
\label{eq:6.1}
\end{equation}
\begin{equation}
\mathcal{D}^{02}
(\scriptstyle 12 \displaystyle)=
-\frac{\delta^{2}\Phi}
{\delta v_{1}^{02}v_{2}^{20}}\,.
\label{eq:6.2}
\end{equation}

To write the equations of motion for the GFs
$\mathcal{D}^{\sigma\bar{\sigma}}$ and $\mathcal{D}^{02}$ we need
the equations of motion for the Bose-like operators:
\begin{equation}
\dot{X}_{1}^{\sigma\bar{\sigma}}=
-(\varepsilon_{\bar{\sigma}}
-\varepsilon_{\sigma})
X_{1}^{\sigma\bar{\sigma}}-
\varPsi_{\alpha^{\prime}}^{\dag}
(\scriptstyle 1 \displaystyle \sigma)
\Im_{\alpha^{\prime}\beta^{\prime}}
t(\scriptstyle 11^{\prime} \displaystyle)
\varPsi_{\beta^{\prime}}
(\scriptstyle 1^{\prime} \displaystyle\bar{\sigma})+
\varPsi_{\alpha^{\prime}}^{\dag}
(\scriptstyle 1^{\prime} \displaystyle\sigma)
t(\scriptstyle 1^{\prime}1 \displaystyle)
\Im_{\alpha^{\prime}\beta^{\prime}}
\varPsi_{\beta^{\prime}}
(\scriptstyle 1 \displaystyle \bar{\sigma}),
\label{eq:6.3}
\end{equation}
\begin{equation}
\dot{X}_{1}^{02}=
-(U-2\mu)X_{1}^{02}
+\sigma^{\prime}(\tau^{x}\varPsi)_{\alpha^{\prime}}
(\scriptstyle 1 \displaystyle \bar{\sigma}^{\prime})
\Im_{\alpha^{\prime}\beta^{\prime}}
t(\scriptstyle 11^{\prime} \displaystyle)
\varPsi_{\beta^{\prime}}
(\scriptstyle 1^{\prime} \displaystyle\sigma^{\prime}).
\label{eq:6.4}
\end{equation}
We see that in the right hand sides of these relations
$\varPsi$-operators occur; therefore in the corresponding
equations of motions for the magnon and the doublon GFs the
$T$-mixed product of $f$- and $b$-operators will appear. They
could be represented as the the variational derivative of the
electronic GF with respect to the fluctuating field
$v^{\sigma\bar{\sigma}}$ in the first case and $v^{02}$ in the
second. One of the important feature of the doublon GF is that it
includes the ``anomalous" electronic GF, composed of the two
operators $\varPsi(\scriptstyle 1 \displaystyle \sigma)$ and
$\varPsi(\scriptstyle 2 \displaystyle \bar{\sigma})$, while the
equation for the magnon GF should include the normal electronic
GF, composed of the operators $\varPsi(\scriptstyle 1
\displaystyle \sigma)$ and $\varPsi^{\dag}(\scriptstyle 2
\displaystyle \bar{\sigma})$. By itself these anomalous GFs are
equal to zero when the fields are absent, however their
derivatives with respect to the fields $v^{\sigma\bar{\sigma}}$
and $v^{02}$ are not equal to zero and determine the contribution
in the equation of motion, caused by the interactions of the
electronic and bosonic degrees of freedom. Now our task is to
determine the equations of motion for the magnon and doublon GFs
and to obtain their approximate solution. This will let us
determine the spectrum of the corresponding collective modes. In
this paper we study only the doublon GF.

Let us derive the equation of motion for the doublon GF
(\ref{eq:2.22}); to this purpose we write the equation of motion
for the mean value of the operator $X_{1}^{02}$:
\begin{equation}
\frac{\partial}{\partial\tau_{1}}
(\!(TX_{1}^{02}e^{-V})\!)=
(\!(T\dot{X}_{1}^{02}e^{-V})\!)-
(\!(T\{X_{1}^{02},V\}e^{-V})\!).
\label{eq:6.5}
\end{equation}
We substitute in it the expression (\ref{eq:6.4})
for $\dot{X}_{1}^{02}$, and also the relation
\[
\{X_{1}^{02},V\}=
-v_{1}^{00}X_{1}^{02}+v_{1}^{22}X_{1}^{02}
+v_{1}^{02}(X_{1}^{00}-X_{1}^{22}).
\]
Then our initial equation could be rewritten in the form:
\begin{align*}
\left(K_{0V}^{02}\right)^{-1}&
(\scriptstyle 11^{\prime}\displaystyle)
(\!(TX_{1^{\prime}}^{02}e^{-V})\!)=
-v_{1}^{02}(\!(T(X_{1}^{22}-X_{1}^{00})e^{-V})\!) \\
+&\sigma^{\prime}
t(\scriptstyle 11^{\prime}\displaystyle)
\Im_{\alpha^{\prime}\beta^{\prime}}
(\!(T\varPsi_{\beta^{\prime}}
(\scriptstyle 1^{\prime}\displaystyle\sigma^{\prime})
(\tau^{x}\varPsi)_{\alpha^{\prime}}
(\scriptstyle 1 \displaystyle\bar{\sigma}^{\prime})
e^{-V})\!),
\end{align*}
where
\begin{equation}
\left(K_{0V}^{02}\right)^{-1}
(\scriptstyle 12 \displaystyle)=
-\left(
\frac{\partial}{\partial\tau_{1}}+U-2\mu+
v_{1}^{22}-v_{1}^{00} \right)\delta_{12}.
\label{eq:6.6}
\end{equation}
Taking into account the definition of the electronic GF
we write its last term in the form
\[
-Z\sigma^{\prime}\,\mathrm{Tr}
\left[\Im
\left(t\mathcal{L}^{12}
\right)
(\scriptstyle 1 \displaystyle \sigma^{\prime},
\scriptstyle 1 \displaystyle \bar{\sigma}^{\prime})
\right],
\]
where
\begin{equation}
\mathcal{L}_{\alpha\beta}^{12}
(\scriptstyle 1 \displaystyle\sigma_{1},
\scriptstyle 2 \displaystyle\sigma_{2})=
-\langle T\varPsi_{\alpha}
(\scriptstyle 1 \displaystyle\sigma_{1})
\varPsi_{\beta}(\scriptstyle 2 \displaystyle\sigma_{2})
\rangle
\label{eq:6.7}
\end{equation}
is the anomalous component of the electronic GF.

The mean values $(\!(\dots)\!)$ of $X$ operators are expressed
through the variational derivative of the functional $\Phi[V]$,
and we come to the final form of the equation for the generating
functional:
\begin{equation}
\left(K_{0V}^{02}\right)^{-1}
(\scriptstyle 11^{\prime}\displaystyle)
\frac{\displaystyle \delta\Phi}
{\displaystyle\delta v_{1^{\prime}}^{02}}=
-v_{1}^{02}
\left(
\frac{\displaystyle \delta\Phi}
{\displaystyle \delta v_{1}^{22}}-
\frac{\displaystyle \delta\Phi}
{\displaystyle\delta v_{1}^{00}}
\right)
+\sigma^{\prime}\,\mathrm{Tr}
\left[
\Im\left(
t\mathcal{L}^{12}
\right)
(\scriptstyle 1 \displaystyle\sigma^{\prime},
\scriptstyle 1 \displaystyle\bar{\sigma}^{\prime})
\right].
\label{eq:6.8}
\end{equation}
In the same way it is possible to write the equation for
$(\!(TX_{1}^{20}e^{-V})\!)$ and reduce it to the form
\begin{equation}
\frac{\delta\Phi}{\delta v_{1^{\prime}}^{20}}
\left(K_{0V}^{02}\right)^{-1}
(\scriptstyle 1^{\prime}1 \displaystyle)
=
-v_{1}^{20}
\left(
\frac{\displaystyle \delta\Phi}
{\displaystyle \delta v_{1}^{22}}-
\frac{\displaystyle \delta\Phi}
{\displaystyle\delta v_{1}^{00}}
\right)
+\sigma^{\prime}\,\mathrm{Tr}
\left[
\Im\left(
\mathcal{L}^{21}t
\right)
(\scriptstyle 1 \displaystyle\bar{\sigma}^{\prime},
\scriptstyle 1 \displaystyle\sigma^{\prime})
\right].
\label{eq:6.9}
\end{equation}
Differentiating now the equation (\ref{eq:6.8}) with respect to
$v_{2}^{20}$, and the equation (\ref{eq:6.9}) with respect to
$v_{2}^{02}$, we come to the pair of conjugate equations for the
doublon GF:
\begin{equation}
\left(K_{0V}^{02}\right)^{-1}
(\scriptstyle 11^{\prime}\displaystyle)
\mathcal{D}^{02}
(\scriptstyle 1^{\prime}2 \displaystyle)=
(1-n_{1})\delta_{12}
-\sigma^{\prime}\frac{\delta}{\delta v_{2}^{20}}
\mathrm{Tr}
\left[
\Im\left(t\mathcal{L}^{12}\right)
(\scriptstyle 1 \displaystyle\sigma^{\prime},
\scriptstyle 1 \displaystyle\bar{\sigma}^{\prime})
\right].
\label{eq:6.10}
\end{equation}
\begin{equation}
\mathcal{D}^{02}
(\scriptstyle 12^{\prime}\displaystyle)
\left(K_{0V}^{02}\right)^{-1}
(\scriptstyle 2^{\prime}2\displaystyle)=
(1-n_{1})\delta_{12}-
\sigma^{\prime}\frac{\delta}{\delta v_{2}^{02}}
\mathrm{Tr}
\left[
\Im\left(\mathcal{L}^{21}t\right)
(\scriptstyle 2\displaystyle\bar{\sigma}^{\prime},
\scriptstyle 2\displaystyle \sigma^{\prime})
\right].
\label{eq:6.11}
\end{equation}
Here we introduced the number of electrons on the site,
$n_{1}=n_{1}^{\sigma}+n_{1}^{\bar{\sigma}}$, where
$n_{1}^{\sigma}=\langle c_{1\sigma}^{\dag}c_{1\sigma}\rangle$. We see
that the exact equations for the doublon GF contain terms with
variational derivatives of anomalous electronic GF with respect to
the fields $v_{1}^{02}$ and $v_{1}^{20}$. To obtain a close
equation for doublon GF we have to calculate these terms by the
same approximate way.

Let us calculate the derivative of off-diagonal (with respect to
the upper spinor indexes) electronic GFs $\mathcal{L}^{12}$ and
$\mathcal{L}^{21}$. We use the multiplicative representation
(\ref{eq:3.18}). In the normal state we could use the expression
for the variational derivative.
\begin{equation}
\frac{\delta \mathcal{L}^{12}
(\scriptstyle 34 \displaystyle)}
{\delta v_{2}^{20}}=
-L^{11}(\scriptstyle 33^{\prime} \displaystyle)
\frac{\delta
\left[
{L}^{-1}
(\scriptstyle 3^{\prime }4^{\prime }\displaystyle)
\right]^{12}}
{\delta v_{2}^{20}}
\mathcal{L}^{22}
(\scriptstyle 4^{\prime }4\displaystyle)
+L^{11}(\scriptstyle 33^{\prime}\displaystyle)
\frac{\delta\Pi^{12}
(\scriptstyle 3^{\prime }4\displaystyle)}
{\delta v_{2}^{20}}
\label{eq:6.12}
\end{equation}
We take the inverse propagator GF $L^{-1}$ in the approximation,
when in the general expression (\ref{eq:3.20}) the term $\Sigma
^{\prime }$ is neglected, and $\Pi$ is taken in the zeroth order
approximation. It is easy to obtain the relations:
\[
\frac{\delta
\left[{L}_{0V}^{-1}
(\scriptstyle 3 \displaystyle \sigma _{3},
\scriptstyle 4 \displaystyle \sigma _{4})
\right] ^{12}}
{\delta v_{2}^{20}}=
\sigma _{3}\delta _{\bar{\sigma}_{3}\sigma _{4}}
\delta_{23}\delta _{34}\tau ^{x},
\]
\[
\frac{\delta \Pi _{0}^{12}
(\scriptstyle 3 \displaystyle\sigma _{3},
\scriptstyle 4 \displaystyle\sigma _{4})}
{\delta v_{2}^{20}}=-
\sigma _{3}\delta _{\bar{\sigma}_{3}\sigma _{4}}
\delta _{34}\frac{\delta^{2}\Phi}
{\delta v_{3}^{02}v_{2}^{20}}i\tau ^{y}.
\]
Then, within the first order approximation with respect to $t$ the
Eq. (\ref{eq:6.10}) is determined by the expression
\begin{align}
\frac{\delta\mathcal{L}^{12}
(\scriptstyle 3\displaystyle\sigma _{3},
\scriptstyle4\sigma _{4}\displaystyle)}
{\delta v_{2}^{20}}&
=-\sigma _{3}\delta _{\bar{\sigma}_{3}\sigma _{4}}
\biggl\{
G^{\sigma_{3}}
(\scriptstyle 32 \displaystyle)
\tau ^{x}
\widetilde{\mathcal{G}}^{\bar{\sigma _{3}}}
(\scriptstyle 24 \displaystyle)-
\label{eq:6.13}
\\
&-G^{\sigma _{3}}
(\scriptstyle 33^{\prime } \displaystyle)
\left[
\delta _{3^{\prime }4}i\tau^{y}-i\tau ^{y}\Im
(\tilde{t}\,\widetilde{\mathcal{G}}^{\bar{\sigma _{3}}})
(\scriptstyle 3^{\prime }4 \displaystyle)
\right]
\mathcal{D}^{02}
(\scriptstyle 3^{\prime }2 \displaystyle)
\biggr\}.
\nonumber
\end{align}
After substituting this relation into the equation
(\ref{eq:6.10}), we represent equation for the doublon GF in the
form
\begin{equation}
\left[
\left(K_{0V}^{02}\right)^{-1}
(\scriptstyle 11^{\prime}\displaystyle)
-\mathcal{M}_{l}^{02}
(\scriptstyle 11^{\prime}\displaystyle)
\right]
\mathcal{D}^{02}
(\scriptstyle 1^{\prime}2 \displaystyle)=
(1-n_{1})\delta _{12}+\mathcal{P}_{l}^{02}
(\scriptstyle 12 \displaystyle),
\label{eq:6.14}
\end{equation}
\begin{equation}
\mathcal{P}_{l}^{02}
(\scriptstyle 12 \displaystyle)=
\mathrm{Tr}
\left[\Im(tG^{\sigma ^{\prime}})
(\scriptstyle 12 \displaystyle)
\tau ^{x}\widetilde{\mathcal{G}}^{\bar{\sigma}^{\prime}}
(\scriptstyle 21 \displaystyle)
\right],
\label{eq:6.15}
\end{equation}
\begin{equation}
\mathcal{M}_{l}^{02}
(\scriptstyle 12 \displaystyle)
=-\mathrm{Tr}
\left[\Im(tG^{\sigma ^{\prime }})
(\scriptstyle 12 \displaystyle)
\left(i\tau ^{y}\delta _{12}-i\tau ^{y}\Im
(\tilde{t}\,\widetilde{\mathcal{G}}^{\bar{\sigma}^{\prime }})
(\scriptstyle 21 \displaystyle)
\right)
\right].
\label{eq:6.16}
\end{equation}
The index $l$ of the terminal and the self-energy part indicates
the ``left" form of the equations for $\mathcal{D}^{02}$. In the
same way starting from the equation (\ref{eq:6.11}), it is
possible to come to the ``right" form of the equation for
$\mathcal{D}^{02}$:
\begin{equation}
\mathcal{D}^{02}
(\scriptstyle 12^{\prime} \displaystyle)
\left[
\left( K_{0V}^{02}\right)^{-1}
(\scriptstyle 2^{\prime}2 \displaystyle)
-\mathcal{M}_{r}^{02}
(\scriptstyle 2^{\prime}2 \displaystyle)
\right]
=(1-n_{1})\delta _{12}
+\mathcal{P}_{r}^{02}
(\scriptstyle  12 \displaystyle),
\label{eq:6.17}
\end{equation}
\begin{equation}
\mathcal{P}_{r}^{02}
(\scriptstyle 12 \displaystyle)=
\mathrm{Tr}
\left[
\Im \widetilde{G}^{\bar{\sigma}^{\prime}}
(\scriptstyle 21 \displaystyle)
\tau ^{x}(\mathcal{G}^{\sigma ^{\prime }}t)
(\scriptstyle 12 \displaystyle)
\right],
\label{eq:6.18}
\end{equation}
\begin{equation}
\mathcal{M}_{r}^{02}
(\scriptstyle 12 \displaystyle)=
\mathrm{Tr}
\left[
\Im \widetilde{G}^{\bar{\sigma}^{\prime}}
(\scriptstyle 21 \displaystyle)
\left(
i\tau ^{y}t_{12}+i\tau ^{y}\Im
(t\mathcal{G}^{\sigma ^{\prime}}t)
(\scriptstyle 12 \displaystyle)
\right)
\right].
\label{eq:6.19}
\end{equation}
To recover the symmetry of the doublon GF let us symmetrize the
equations (\ref{eq:6.14}) and (\ref{eq:6.17}) making their sum.
Then, the doublon GF is equal to
\begin{equation}
\mathcal{D}^{02}(q)=
\frac{(1-n)+\mathcal{P}^{02}(q)}
{i\omega _{n}-(U-2\mu)-\mathcal{M}^{02}(q)},
\label{eq:6.20}
\end{equation}
where
\[
\mathcal{M}^{02}(q)=
\frac{\displaystyle1}{\displaystyle2}\,
[\mathcal{M}_{l}^{02}(q)+\mathcal{M}_{r}^{02}(q)],
\qquad
\mathcal{P}^{02}(q)=
\frac{\displaystyle1}{\displaystyle2}\,
[\mathcal{P}_{l}^{02}(q)+\mathcal{P}_{r}^{02}(q)].
\]
The self-energy and the terminal part are equal to
\begin{align}
\mathcal{M}^{02}(q) =-\frac{\displaystyle 1}{\displaystyle 2}
\sum\limits_{k\sigma} \varepsilon(\bm{k}) \biggl\{
&\mathrm{Tr}[\Im\,G^{\sigma}(k)\,i\tau^{y}]-
\mathrm{Tr}[\Im\,\widetilde{G}^{\sigma}(k-q)\,i\tau^{y}] \biggr\}
\nonumber
\\
+\frac{\displaystyle 1}{\displaystyle 2} \sum\limits_{k\sigma}
&\varepsilon(\bm{k}) \biggl\{\varepsilon(\bm{k-q})
\mathrm{Tr}[\Im\,G^{\sigma}(k)\,i\tau^{y}\,
\Im\,\widetilde{\mathcal{G}}^{\bar{\sigma}}(k-q)] \label{eq:6.21}
\\
+& \varepsilon(\bm{k})
\mathrm{Tr}[i\tau^{y}\,\Im\,\mathcal{G}^{\sigma}(k)\,
\Im\,\widetilde{G}^{\bar{\sigma}}(k-q)]
\biggr\},
\nonumber
\end{align}
\begin{equation}
\mathcal{P}^{20}(q)=
\frac{\displaystyle 1}{\displaystyle 2}
\sum\limits_{k\sigma}
\varepsilon(\bm{k})
\biggl\{
\mathrm{Tr}[\Im\,G^{\sigma}(k)\,\tau^{x}\,
\widetilde{\mathcal{G}}^{\bar{\sigma}}(k-q)]+
\mathrm{Tr}[\Im\,\widetilde{G}^{\sigma}(k-q)\,\tau^{x}\,
\mathcal{G}^{\bar{\sigma}}(k)]
\biggr\}.
\label{eq:6.22}
\end{equation}
In the same way we can calculate the doublon GF
$\mathcal{D}_{12}^{20}$. It is possible to represent the result of
the computation in the form
\begin{equation}
\mathcal{D}^{20}(q)=
\frac{-(1-n)+\mathcal{P}^{20}(q)}
{i\omega _{n}+(U-2\mu )-\mathcal{M}^{20}(q)},
\label{eq:6.23}
\end{equation}
where the values $\mathcal{P}^{20}(q)$ and $\mathcal{M}^{20}(q)$ are
expressed through $\mathcal{P}^{02}(q)$ and $\mathcal{M}^{02}(q)$:
\begin{equation}
\begin{array}{c}
\mathcal{P}^{20}(q)=-\mathcal{P}^{02}(-q),\\
\\
\mathcal{M}^{20}(q)=-\mathcal{M}^{02}(-q).
\end{array}
\label{eq:6.24}
\end{equation}
Thus we see that the condition of symmetry is fulfilled
\begin{equation}
\mathcal{D}^{20}(q)=\mathcal{D}^{02}(-q),
\label{eq:6.25}
\end{equation}
or $\mathcal{D}_{12}^{20}=\mathcal{D}_{21}^{02}$ in the coordinate
space, which follows directly from the definition (\ref{eq:6.2})
for the doublon GF.

After the computation of the trace in the expressions
(\ref{eq:6.21}) and (\ref{eq:6.22}), we can represent them in the
form:
\begin{align}
&\mathcal{M}^{02}(q)= -\frac{\displaystyle 1}{\displaystyle 2}
\sum\limits_{k\sigma} \varepsilon(\bm{k})
[g^{\sigma}(k)-\widetilde{g}^{\sigma}(k-q)] \label{eq:6.26}
\\
+\frac{\displaystyle 1}{\displaystyle 2}
\sum\limits_{k\sigma}
\varepsilon(\bm{k})&
\biggl[
\varepsilon(\bm{k-q})g^{\sigma}(k)
\sum\limits_{\alpha\beta}
\widetilde{\mathcal{G}}_{\alpha\beta}^{\bar{\sigma}}(k-q)+
\varepsilon(\bm{k})\widetilde{g}^{\sigma}(k-q)
\sum\limits_{\alpha\beta}
\mathcal{G}_{\alpha\beta}^{\bar{\sigma}}(k)\biggr].
\nonumber
\end{align}
\begin{align}
\mathcal{P}^{02}(q)= -\frac{\displaystyle 1}{\displaystyle 2}
\sum\limits_{k\sigma}\varepsilon(\bm{k}) \biggl\{
&[G_{11}^{\sigma}(k)+G_{21}^{\sigma}(k)]\,
[\mathcal{G}_{12}^{\bar{\sigma}}(-k+q)+
\mathcal{G}_{22}^{\bar{\sigma}}(-k+q)] \label{eq:6.27}
\\
+&[G_{12}^{\sigma}(k)+G_{22}^{\sigma}(k)]\,
[\mathcal{G}_{11}^{\bar{\sigma}}(-k+q)+
\mathcal{G}_{21}^{\bar{\sigma}}(-k+q)]
\nonumber\\
+&[G_{11}^{\sigma}(-k+q)+G_{12}^{\sigma}(-k+q)]\,
[\mathcal{G}_{21}^{\bar{\sigma}}(k)+
\mathcal{G}_{22}^{\bar{\sigma}}(k)]
\nonumber\\
+&[G_{21}^{\sigma}(-k+q)+G_{22}^{\sigma}(-k+q)]\,
[\mathcal{G}_{11}^{\bar{\sigma}}(k)+
\mathcal{G}_{12}^{\bar{\sigma}}(k)]
\biggr\}.
\nonumber
\end{align}
Now we calculate the expression (\ref{eq:6.26}) in the mean field
approximation for the electronic GF. Substituting here the formula
(\ref{eq:5.3}) and summing over frequencies, we write the result
as a sum of two contributions of the first and second order with
respect to $t$:
\[
\mathcal{M}^{02}(q)=\mathcal{M}_{1}^{02}(\bm{q})+\mathcal{M}_{2}^{02}(q),
\]
where
\begin{equation}
\mathcal{M}_{1}^{02}(\bm{q})=
-\frac{U}{2}\sum\limits_{\bm{k}\sigma}
\frac{\varepsilon(\bm{k})+\varepsilon(\bm{k-q})}
{2Q^{\sigma}(\bm{k})}
\biggl[
f\left[E_{\scriptscriptstyle 1\displaystyle}^{\sigma}(\bm{k})
\right]-
f\left[E_{\scriptscriptstyle 2\displaystyle}^{\sigma}(\bm{k})
\right]
\biggr],
\label{eq:6.28}
\end{equation}
\begin{align}
\mathcal{M}_{2}^{02}(q)= -\frac{\displaystyle
1}{\displaystyle2}\sum\limits_{\bm{k}\sigma}&
\varepsilon(\bm{k-q}) \frac{\displaystyle
\varepsilon(\bm{k})+\varepsilon(\bm{k-q})} {\displaystyle
2Q^{\sigma}(\bm{k})}
\sum\limits_{nm}C_{nm}^{\sigma}(\bm{k},\bm{k-q}) \nonumber
\\
\times&
\frac{\displaystyle 1 -
f\left[E_{\scriptscriptstyle n \displaystyle}^{\sigma}(\bm{k})
\right]-
f\left[E_{\scriptscriptstyle m \displaystyle}^{\bar{\sigma}}(\bm{k-q})
\right]}{i\omega_{\nu}-
E_{\scriptscriptstyle n \displaystyle}^{\sigma}(\bm{k})-
E_{\scriptscriptstyle m \displaystyle}^{\bar{\sigma}}(\bm{k-q})}.
\label{eq:6.29}
\end{align}
where $(n,m=1,2)$
\begin{align}
C_{nm}^{\sigma}(\bm{k},\bm{k-q})=&
[(\mathcal{A}_{n}^{\sigma})_{11}-
(\mathcal{A}_{n}^{\sigma})_{22}](\bm{k})
\biggl\{\bigl[(\mathcal{A}_{m}^{\bar{\sigma}})_{11}+
(\mathcal{A}_{m}^{\bar{\sigma}})_{21}\bigl](\bm{k})
\bigl(\Lambda_{11}^{\bar{\sigma}}+
\Lambda_{12}^{\bar{\sigma}}\bigl)
+
\nonumber
\\
+&\bigl[(\mathcal{A}_{m}^{\bar{\sigma}})_{12}+
(\mathcal{A}_{m}^{\bar{\sigma}})_{22}\bigl](\bm{k})
\bigl(\Lambda_{21}^{\bar{\sigma}}+
\Lambda_{22}^{\bar{\sigma}}\bigl)
\biggl\}
\label{eq:6.30}
\end{align}
and here $(\mathcal{P}_{m}^{\sigma})$, $\Lambda^{\sigma}$ are
determined by formulas (\ref{eq:5.13}) and (\ref{eq:4.5})
Remarkable is the fact that expressions $\mathcal{M}_{1}$ and
$\mathcal{M}_{2}$ vanish at wave vector $\bm{Q}=(\pi,\pi,\dots)$.
Because $\mathcal{M}^{02}(q)$ is nothing but the self-energy of a
doublon, we see from Eq. (\ref{eq:6.23}), that at half-filling,
when $U-2\mu=0$, a doublon is a soft mode in the vicinity of the
point $(\pi,\pi,\dots)$. This observation pushes to study its
dispersion law and attenuation.

We postpone the study of doublons at arbitrary electron
concentration and fix ourselves on the case $n=1$. We are limited
now to the hydrodynamical regime.

\section{Dynamical fluctuations in the hydrodynamical regime}
It is well known that collective modes in a disordered
(symmetrical) phase in the hydrodynamical regime are ruled by the
conservation laws \cite{fors}. Thus the spin GF
$\mathcal{D}^{\sigma\bar{\sigma}}$ should be determined by the
total spin conservation law, while the pseudospin GF
$\mathcal{D}^{02}$ is determined by the pseudospin conservation
law \cite{ya}\cite{shen}\cite{bruks}\cite{ave6}. The three
pseudospin components
\begin{equation}
P^{+}=\sum\limits_{i}e^{i\bm{Q}\cdot
\bm{R}_{i}}c_{i\uparrow}^{\dag}c_{i\downarrow}^{\dag}, \quad
P^{-}=\sum\limits_{i}e^{i\bm{Q}\cdot
\bm{R}_{i}}c_{i\uparrow}c_{i\downarrow}, \quad
P^{z}=\sum\limits_{i}\frac{1}{2}(n_{i}-1) \label{eq:7.1}
\end{equation}
with $\bm{Q}=(\pi\,\pi\,\dots)$ obey permutation relations
\begin{equation}
[P^{+},\mathcal{H}]=(2\mu-U)P^{+},\quad
[P^{-},\mathcal{H}]=-(2\mu-U)P^{-},\quad
[P^{z},\mathcal{H}]=0,
\label{eq:7.2}
\end{equation}
from which it is clear that at half filling $(n=1)$ all pseudospin
components are conserved. This leads to the diffusion form of the
pseudospin (doublon) susceptibility, which is the retarned doublon
GF $\chi^{02}(\bm{q},\omega)$. According to Kubo-Mori theory, this
susceptibility is expressed through the memory function
$\mathcal{M}^{02}(\bm{q},\omega)$ by the relation
\begin{equation}
\chi^{02}(\bm{q},\omega)=
\langle\!\langle X_{i}^{02}\bigl|X_{j}^{20}\rangle\!\rangle
_{\bm{q},\omega}=
\frac{M^{02}(\bm{q},\omega)}
{\omega-\frac{\displaystyle M^{02}(\bm{q},\omega)}
{\displaystyle \chi^{02}_{\bm{q}}}},
\label{eq:7.3}
\end{equation}
where we introduce a notation for the static susceptibility,
$\chi^{02}_{\bm{q}}\equiv\chi^{02}(\bm{q},0)$.

On the other hand the memory function is
expressed through the irreducible retarded GF of pseudospin
currents (see \cite{juk}):
\begin{equation}
M^{02}(\bm{q},\omega)=
-\sum\limits_{ij}e^{-i\bm{q}\bm{R}_{ij}}
\int\limits_{-\infty}\limits^{+\infty}
\frac{d\omega'}{\pi}\,
\frac{Im \langle\!\langle
i\dot{X}_{i}^{02}\bigl|
-i\dot{X}_{j}^{20}\rangle\!\rangle_{\omega'}^{irr}}
{\omega'(\omega-\omega'+i\delta)}.
\label{eq:7.4}
\end{equation}
Here $\dot{X}_{i}^{02}$ means the time derivative of operator
$X_{i}^{02}$:
\begin{equation}
i\dot{X}_{i}^{02}=[X_{i}^{02},\mathcal{H}]=
(U-2\mu)X_{i}^{02}-\sigma^{\prime}(\tau^{x}\varPsi)_{\alpha^{\prime}}
(i\bar{\sigma}^{\prime})(\hat{t}\varPsi)_{\alpha^{\prime}}(i\sigma^{\prime}).
\label{eq:7.5}
\end{equation}

Further to this, we consider the half-filling case, when
$U-2\mu=0$. Then
\begin{align}
&\langle\!\langle i\dot{X}_{i}^{02}(t)\bigl|-i\dot{X}_{j}^{20}(0)
\rangle\!\rangle^{irr} \label{eq:7.6}
\\
=\langle\!\langle
(\tau^{x}\varPsi)_{\alpha^{\prime}}(i\bar{\sigma}^{\prime},t)
&(\hat{t}\varPsi)_{\alpha^{\prime}}(i\sigma^{\prime},t)\bigl|
(\varPsi^{+}\hat{t})_{\beta^{\prime}}(j\sigma^{\prime})
(\varPsi^{+}\tau^{x})_{\beta^{\prime}}(j\bar{\sigma}^{\prime})
\rangle\!\rangle^{irr}\nonumber
\\
-\langle\!\langle
(\tau^{x}\varPsi)_{\alpha^{\prime}}(i\bar{\sigma}^{\prime},t)
&(\hat{t}\varPsi)_{\alpha^{\prime}}(i\sigma^{\prime},t)\bigl|
(\varPsi^{+}\hat{t})_{\beta^{\prime}}(j\bar{\sigma}^{\prime})
(\varPsi^{+}\tau^{x})_{\beta^{\prime}}(j\sigma^{\prime})
\rangle\!\rangle^{irr}.\nonumber
\end{align}
Now we use the approximation of interacting modes, well known in
the relaxation theory, by Mori \cite{mori}: the two-particle
electron correlations in expression (\ref{eq:7.6}) are decomposed
into pair correlators and then expressed through the imaginary
parts of the retarded electron GFs. As a result we come to the
following expression, determining the memory function:
\begin{align}
M^{02}(\bm{q},\omega)=2\sum\limits_{\bm{k}}
&\bigl[\varepsilon(\bm{k})+\varepsilon(\bm{k}-\bm{q})\bigr]^{2}
\int d\omega^{\prime}\int d\omega_{1} \bigl[
f(\omega_{1}-\omega')-f(\omega_{1}) \bigr] \label{eq:7.7}
\\
&\times
\frac{\displaystyle
tr\bigl\{[Im\mathcal{G}^{\bar{\sigma}^{\prime}}
(\bm{q}-\bm{k},\omega^{\prime}-\omega_{1})]
[Im(\Im\mathcal{G}^{T\sigma^{\prime}}(\bm{k},\omega_{1})\Im)]
\bigr\}}
{\omega^{\prime}(\omega-\omega^{\prime}+i\delta)}\nonumber.
\end{align}
Here $\mathcal{G}^{T\sigma}$ is the transposed matrix
$\mathcal{G}^{\sigma}$. The quantity
$\mathcal{G}^{\sigma}(\bm{k},\omega)$ is the retarded electron GF.
It can be obtained from our Matsubara GFs by analytical
continuation from discrete imaginary frequencies into real ones:
$i\omega_{n}\to\omega+i\delta$.

Expression (\ref{eq:7.7}) is similar to those obtained in the
interacting modes approximation for other dynamical
susceptibilities. For example, the spin susceptibility is:
\begin{equation}
\chi^{\sigma\bar{\sigma}}(\bm{q},\omega)=
\langle\!\langle
X_{i}^{\sigma\bar{\sigma}}\bigl|
X_{j}^{\bar{\sigma}\sigma}
\rangle\!\rangle_{\bm{q},\omega}=
-\frac{M^{\sigma\bar{\sigma}}(\bm{q},\omega)}
{\omega-\frac{\displaystyle M^{\sigma\bar{\sigma}}(\bm{q},\omega)}
{\displaystyle \chi_{\bm{q}}^{\sigma\bar{\sigma}}}}.
\label{eq:7.8}
\end{equation}
By similar decoupling of the irreducible GFs of the currents we
obtain:
\begin{align}
M^{\sigma\bar{\sigma}}(\bm{q},\omega)= &4\sum\limits_{\bm{k}}
\bigl[\varepsilon(\bm{k})-\varepsilon(\bm{k}-\bm{q})\bigr]^{2}
\int d\omega^{\prime}\int d\omega_{1}
\bigl[f(\omega_{1}-\omega^{\prime})-f(\omega_{1})\bigr]
\label{eq:7.9}
\\
&\times
\frac{\displaystyle
tr\bigl\{[Im\mathcal{G}^{\sigma}(\bm{k}-\bm{q},\omega_{1}-\omega^{\prime})]
[Im(\Im\mathcal{G}^{\bar{\sigma}}\Im)(\bm{k},\omega]
\bigr\}}
{\displaystyle\omega^{\prime}(\omega-\omega^{\prime}+i\delta)}.\nonumber
\end{align}

It is remarkable that the memory GF for the spin susceptibility
vanishes at $\bm{q}=0$, while for the doublon susceptibility it
vanishes at $\bm{q}=\bm{Q}$. This difference originates from the
total spin conservation law (Fourier component of the spin density
at $\bm{q}=0$), while the component of the pseudospin density is
conserved at $\bm{q}=\bm{Q}$ and only for half-filling. There is
another important difference in expressions (\ref{eq:7.7}) and
(\ref{eq:7.9}). Arguments of the electron GFs appear in a
different way in these expressions. This reflects the fact that
the spin collective mode is formed through excitations of a
particle and a hole, while the pseudospin collective mode
(doublone) is formed through excitations of two particles (or two
holes).

Consider now the hydrodynamical limit corresponding to small
frequencies $\omega$ and small wave point $\bm{q}$ (for expression
(\ref{eq:7.7})). In the hydrodynamical limit $\omega\ll vq$, where
$v$ is a characteristic electron velocity on the Fermi surface.
Under these conditions from Eqs. (\ref{eq:7.7}) and (\ref{eq:7.9})
the asymptotic expressions follow:
\begin{equation}
Im\mathcal{M}^{02}(\bm{q},\omega)=-D^{02}p^{2},
\qquad
Re\mathcal{M}^{02}(\bm{q},\omega)=0,
\label{eq:7.10}
\end{equation}
\begin{equation}
Im\mathcal{M}^{\sigma\bar{\sigma}}(\bm{q},\omega)=
-D^{\sigma\bar{\sigma}}q^{2},
\qquad
Re\mathcal{M}^{\sigma\bar{\sigma}}(\bm{q},\omega)=0,
\label{eq:7.11}
\end{equation}
where the coefficients of spin and pseudospin stiffness are equal
\begin{equation}
D^{02}=2\pi\sum\limits_{\bm{k}}
\bigl(v(\bm{k})\bm{e}\bigr)^{2}
\int d\omega_{1}f^{\prime}(\omega_{1})
tr\biggl\{
Im\mathcal{G}^{\bar{\sigma}^{\prime}}(-\bm{k},-\omega_{1})
Im\bigl[\Im
\mathcal{G}^{T\sigma^{\prime}}(\bm{k},\omega_{1})
\Im\bigr]
\biggr\},
\label{eq:7.12}
\end{equation}
\begin{equation}
D^{\sigma\bar{\sigma}}=4\pi\sum\limits_{\bm{k}}
\bigl(v(\bm{k})\bm{e}\bigr)^{2}
\int d\omega_{1}f^{\prime}(\omega_{1})
tr\biggl\{
Im\mathcal{G}^{\sigma}(\bm{k},\omega_{1})
Im\bigl[\Im
\mathcal{G}^{\bar{\sigma}}(\bm{k},\omega_{1})
\Im\bigr]
\biggr\}.
\label{eq:7.13}
\end{equation}
Here $\bm{e}$ is the unit wave vector, and $f^{\prime}(\omega)$ is
the derivative of the Fermi function.

Expression (\ref{eq:7.13}) is valid at arbitrary $U$; in the case
of $U \gg W$ it is consistent with the result of \cite{juk} for
the $tJ$-model.

Notice that if we use the electron GF in the mean field
approximation (without attenuation of quasiparticles) both
expressions (\ref{eq:7.12}) and (\ref{eq:7.13}) vanish. It is easy
to show that if the attenuation of quasiparticles $\gamma$ obeys
the condition $\gamma\gg vq$, both expressions become finite. In
the general case expressions (\ref{eq:7.7}) and (\ref{eq:7.9}) for
the memory function give in the hydrodynamical limit correct
asymptotic values, therefore the susceptibilities have the
diffusion form, which is
\begin{equation}
\frac{1}{\omega}Im\chi(\bm{q},\omega)=
\chi_{\bm{q}}\frac{\displaystyle \widetilde{D}q^{2}}
{\omega^{2}+\bigl(\widetilde{D}q^{2}\bigr)^{2}},
\label{eq:7.14}
\end{equation}
where $\widetilde{D}=D/\chi_{\bm{q}}$.


\section{Conclusions}
We have applied the GFA to investigate the Hubbard model in the
$X$ operator representation. This means that we discussed the case
of sufficiently strong electronic correlations $U>W$. We have
derived the exact equation for the electronic GF in terms of the
variational derivatives with respect to the fluctuating fields
$v^{\sigma\sigma^{\prime}}$, $v^{\sigma\bar{\sigma}}$, $v^{20}$
coupled with the spin and charge densities. The electronic GFs
represent generally an 8x8 matrix with respect to the three
discrete indexes $\sigma$, $\alpha$, $\nu$. In the matrix
representation the equation has the same structure with the GF for
the Hubbard model in the limit $U\rightarrow\infty$, for the $tJ$-
and $sd$-models and for the GFs of the transverse spin components
in the Heisenberg model as well.

The electronic GF $\mathcal{G}$ has a multiplicative character in
the sense that it is expressed by a product of two quantities,
$\mathcal{G}=G\Lambda$, where $G$ is the propagator satisfying the
Dyson equation with the self energy $\Sigma$, and $\Lambda$ is the
terminal part. From the equation for $\mathcal{G}$, a pair of
equations with variational derivatives for $\Sigma$ and $\Lambda$
follow. Their iteration generates a power series in the parameter
$W/U$. This corresponds to the perturbation theory close to the
atomic limit. The iteration corrections of the first two orders allow to
formulate a mean field approximation essentially equivalent to
that of COM.

Taking the electronic GF in the mean field approximation we
derived an equation for the doublon GF. The properties of the
poles of the doublon GF depends substantially on the electronic
concentration $n$. For $n<1$ there is a pole which has a real part
$U-2\mu>0$, corresponding to the activated mode with the quadratic
dispersion law. For $n\rightarrow1$, $U-2\mu\rightarrow0$. The
investigation of the special case $n=1$ reveals that a soft mode
with $\bm{Q}=(\pi,\pi,\dots)$ may exist. However at
$\bm{Q}=(\pi,\pi,\dots)$ the paramagnetic phase of the Hubbard
model has an instability to antiferromagnetic ordering. It means
that two possible instabilities -- doublon and magnon ones --
should compete, and a final result concurring a type of ordering
at half filling demands a farther investigations. It will be a
subject of next study.

The other direction is to investigate magnetically ordered states.
We should go out of the scope of the mean field approximation and
take into account the second order correction $\Sigma_{2}$,
including the interaction of electrons with magnons. The
preliminary analysis reveals that it contains a singular
Kondo-like term $\sim\ln{\mid\omega-E_F\mid}$, which, as it has
been pointed out in the works \cite{he}\cite{zar}, leads to a stable
ferromagnetism. After the extraction of the relevant term in the
second order correction we could write a more exact equation for
the magnon GF and calculate the spin-wave spectrum. All of this
will form the subject of a next paper.

\section*{Acknowledgments}

One of the authors (Yu.A. I.) would like to thank Prof. F. Mancini
for hospitality during his stay in the Dipartimento di Fisica
``E.R. Caianiello", Universit\`{a} degli Studi di Salerno, where
this paper has been started. Yu.A. Izuymov, N.I. Chaschin and D.
S. Alexeev thank the Russian Foundation of Support of Scientific
School, grant NS--747.2003.2.

\section*{Appendix A}
\subsection*{Equation of motion for an arbitrary Green's function}
Consider an average of an arbitrary $T$-product of the operators
$A_1$, $B_2$, $C_3$, $D_4$,... (it could be $X$ operators, spin
operators or other), taken in the Heiseberg representation:

\begin{equation}
A_{1}\equiv\mathrm{e}^{\mathcal{H}\tau}
A_{i_{1}}\mathrm{e}^{-\mathcal{H}\tau},
\qquad 1=\{i_{1},\tau_{1}\},
\tag{A.1}
\label{eq:A.1}
\end{equation}
and so on. Then the following identity is valid:
\begin{align}
\frac{\partial}{\partial\tau_{1}}
(\!(TA_{1}B_{2}C_{3}D_{4}\dots\mathrm{e}^{-V})\!) &=
(\!(T\dot{A}_{1}B_{2}C_{3}D_{4}\dots\mathrm{e}^{-V})\!)\nonumber
\\
+(\!(T\{A_{1},B_{2}\}C_{3}D_{4}\dots\mathrm{e}^{-V})\!)&+
(\!(T\{A_{1},C_{3}\}B_{2}D_{4}\dots\mathrm{e}^{-V})\!)+\dots
\tag{A.2}
\label{eq:A.2} \\
-(\!(T\{A_{1},V\}&B_{2}C_{3}D_{4}\dots\mathrm{e}^{-V})\!).  \nonumber
\end{align}
Here the average over the Gibbs statistical ensemble is denoted by
\begin{equation}
(\!(T...\mathrm{e}^{-V})\!)
\equiv\text{Tr}\{\mathrm{e}^{-\beta\mathcal{H}}T...\mathrm{e}^{-V}\}.
\tag{A.3}
\label{eq:A.3}
\end{equation}
The relation (\ref{eq:A.2}) represents the result of the
differentiation of the initial average with respect to the time
$\tau_1$ ascribed to the operator $A_1$. In the right hand side of
the relation (\ref{eq:A.1}) $\dot{A}_1$ represents the time
derivative
\begin{equation}
\dot{A}_{1}=-[A_{1},\mathcal{H}].
\tag{A.4}
\label{eq:A.4}
\end{equation}
The curl brackets of the kind $\{A_1,B_2\}$ mean
\begin{equation}
\{A_{1},B_{2}\}=\delta(\tau_{1}-\tau_{2})[A_{i_{1}},B_{i_{2}}]_{\pm}
(\tau_{1}),
\tag{A.5}
\label{eq:A.5}
\end{equation}
where $[...,...]_{\pm}$ is an anticommutator or a commutator
depending on the kind of the $A$ and $B$ operators. The signs of
the terms in the second line of Eq.(\ref{eq:A.2}) are determined
by the signs of the transpositions of $f$-operators in the
$T$-product from their original place to the second one in the
energy term. In the last term of Eq.(\ref{eq:A.2}) the expression
$\{A_{1},V\}$ is a commutator because the operator $V$ is implied
to be boson-like. Formally this term and the first one could be
merged and $\mathcal{H}+V$ may be considered in a sense as the
Hamiltonian of a system immersed in fluctuating fields.

The identity (\ref{eq:A.2}) could be proved by differentiating the
$T$-product expressed through the $\theta(\tau-\tau^{\prime})$
functions and the expansion of the exponent $e^{-V}$ in a series,
with subsequent recollection of all the terms back to the
exponent. This identity serves as a basis for the derivation of
the GFs in the fluctuating fields, defined as
\begin{equation}
\langle TA_1B_2\dots\mathrm{e}^{-V} \rangle_V= \frac{(\!(TA_{1}B_{2}\dots
\mathrm{e}^{-V})\!)}{(\!(T\mathrm{e}^{-V})\!)},
\tag{A.6}
\label{eq:A.6}
\end{equation}
where $(\!(T\mathrm{e}^{-V})\!)=Z[V]$ is the generating
functional.


\section*{Appendix B}
\subsection*{Expansion of self-energy and terminal part of the
electronic Green's function} The normal GF
$\mathcal{G}(\scriptstyle 12 \displaystyle)$ can be represented in
a multiplicative form similar to that of (\ref{eq:3.18}) for the
whole GF, namely
\begin{equation}
\mathcal{G}
(\scriptstyle \textbf{12}\displaystyle)
=
G
(\scriptstyle \textbf{11}^{\prime}\displaystyle)
\mathcal{G}
(\scriptstyle \textbf{1}^{\prime}\textbf{2}\,\displaystyle)
\tag{B.1}
\label{eq:B.1}
\end{equation}
where bold indexes mean $\scriptstyle\textbf{1}\displaystyle=
\{\scriptstyle 1 \displaystyle , \alpha , \sigma_{1}\}$. Thus
quantities $\mathcal{G}$, $G$, $\Lambda$ and $\Sigma^{\prime}$
(determined by Dyson equation (\ref{eq:4.3})) are $2\times2$
matrices with respect to the spinor index $\alpha$.

We find the equations for $\Lambda$ and $\Sigma$ from the general
matrix equations (\ref{eq:3.21}) and (\ref{eq:3.22}) by writing
down the equation for the matrix element $\Sigma^{\prime\,11}$:
\begin{align}
&\Sigma^{\prime\,11} (\scriptstyle\textbf{12}\displaystyle)
\tag{B.2} \label{eq:B.2}
\\
= &-\left[ (\hat{t}L^{11})
(\scriptstyle\textbf{4}^{\prime}\textbf{3}^{\prime}\displaystyle)
\hat{A}^{11}
(\scriptstyle\textbf{1}\textbf{4}^{\prime}\displaystyle)-
(\hat{\tilde{t}}L^{21})
(\scriptstyle\textbf{4}^{\prime}\textbf{3}^{\prime}\displaystyle)
\hat{A}^{12}
(\scriptstyle\textbf{1}\textbf{4}^{\prime}\displaystyle)
\right]\centerdot \left[ \left(L_{0V}^{-1}\right)^{11}
(\scriptstyle\textbf{3}^{\prime}\textbf{2}\displaystyle)-
\Sigma^{\prime 11}
(\scriptstyle\textbf{3}^{\prime}\textbf{2}\displaystyle) \right]
\nonumber\\
&-\left[
(\hat{t}L^{12})
(\scriptstyle\textbf{4}^{\prime}\textbf{3}^{\prime}\displaystyle)
\hat{A}^{11}
(\scriptstyle\textbf{1}\textbf{4}^{\prime}\displaystyle)-
(\hat{\tilde{t}}L^{22})
(\scriptstyle\textbf{4}^{\prime}\textbf{3}^{\prime}\displaystyle)
\hat{A}^{12}
(\scriptstyle\textbf{1}\textbf{4}^{\prime}\displaystyle)
\right]\centerdot
\left[
\left(L_{0V}^{-1}\right)^{21}
(\scriptstyle\textbf{3}^{\prime}\textbf{2}\displaystyle)-
\Sigma^{\prime 21}
(\scriptstyle\textbf{3}^{\prime}\textbf{2}\displaystyle)
\right]
\nonumber
\end{align}

\begin{align}
&\Sigma^{\prime\,21} (\scriptstyle\textbf{12}\displaystyle)
\tag{B.3} \label{eq:B.3}
\\
= &-\left[ (\hat{t}L^{11})
(\scriptstyle\textbf{4}^{\prime}\textbf{3}^{\prime}\displaystyle)
\hat{A}^{21}
(\scriptstyle\textbf{1}\textbf{4}^{\prime}\displaystyle)-
(\hat{\tilde{t}}L^{21})
(\scriptstyle\textbf{4}^{\prime}\textbf{3}^{\prime}\displaystyle)
\hat{A}^{22}
(\scriptstyle\textbf{1}\textbf{4}^{\prime}\displaystyle)
\right]\centerdot \left[ \left( L_{0V}^{-1} \right)^{11}
(\scriptstyle\textbf{3}^{\prime}\textbf{2}\displaystyle)-
\Sigma^{\prime 11}
(\scriptstyle\textbf{3}^{\prime}\textbf{2}\displaystyle) \right]
\nonumber\\
&-\left[
(\hat{t}L^{12})
(\scriptstyle\textbf{4}^{\prime}\textbf{3}^{\prime}\displaystyle)
\hat{A}^{21}
(\scriptstyle\textbf{1}\textbf{4}^{\prime}\displaystyle)-
(\hat{\tilde{t}}L^{22})
(\scriptstyle\textbf{4}^{\prime}\textbf{3}^{\prime}\displaystyle)
\hat{A}^{22}
(\scriptstyle\textbf{1}\textbf{4}^{\prime}\displaystyle)
\right]\centerdot
\left[
\left(
L_{0V}^{-1}
\right)^{21}
(\scriptstyle\textbf{3}^{\prime}\textbf{2}\displaystyle)-
\Sigma^{\prime 21}
(\scriptstyle\textbf{3}^{\prime}\textbf{2}\displaystyle)
\right].
\nonumber
\end{align}
For the normal phase, matrix elements $L^{12}=L^{21}=0$, however
the derivatives of them with respect to the fields $v^{02}$ and
$v^{20}$ should not vanish, therefore in equations (\ref{eq:B.2})
and (\ref{eq:B.3}) we must keep such derivatives.

In the first order in $\hat{t}$ equations (\ref{eq:B.2}) and
(\ref{eq:B.3}) lead respectively to an equation for the
self-energy of the normal GF ($\Sigma^{\prime
11}\equiv\Sigma^{\prime}$). Matrix operators $\hat{A}^{11}$ and
$\hat{A}^{12}$ include the derivatives with respect to fluctuating
fields which act on $(L_{0V}^{-1})^{11}$ and $(L_{0V}^{-1})^{12}$,
and therefore we come to the expression:
\begin{align}
&\Sigma_{1}^{\prime} (\scriptstyle 1 \displaystyle \sigma,
\scriptstyle 2 \displaystyle \sigma) \equiv
\Sigma_{1}^{\prime\sigma} (\scriptstyle 12 \displaystyle)
\tag{B.4}
\label{eq:B.4}\\
&=-\delta_{12} \left[ (\tau^{z}\hat{t}G) (\scriptstyle 1
\displaystyle \bar{\sigma}, \scriptstyle 2 \displaystyle
\bar{\sigma})+ (i\tau^{y}\hat{\tilde{t}}\widetilde{G}\tau^{x})
(\scriptstyle 1 \displaystyle \bar{\sigma}, \scriptstyle 2
\displaystyle \bar{\sigma}) \right] \nonumber
\\
&=-\delta_{12}
\biggl[
(tG_{11}^{\bar{\sigma}})
(\scriptstyle 11 \displaystyle)-
(tG_{22}^{\bar{\sigma}})
(\scriptstyle 11 \displaystyle)
\biggr]
\begin{pmatrix}
1 & -1\\
-1 & 1
\end{pmatrix}.
\nonumber
\end{align}
Here in the last line we have used a more concise definition of
$G=G^{\sigma}$, and also made use of the expression for the
transposed matrix
\begin{equation}
\widetilde{G}
(\scriptstyle 12 \displaystyle)=
-G(\scriptstyle 21 \displaystyle).
\tag{B.5}
\label{eq:B.5}
\end{equation}
The calculation of second order contribution in the uncutable part
$\Sigma^{\prime}_{2}$ demands much more efforts, but it involves
nothing else then standard and straightforward iterations of
equations (\ref{eq:B.2}) and (\ref{eq:B.3}). We present the final
result:
\begin{equation}
\Sigma_{2}^{\prime\sigma}
(\scriptstyle 12 \displaystyle)=
-\sum\limits_{\sigma^{\prime}}
(tg^{\sigma^{\prime}})
(\scriptstyle 12 \displaystyle)
\begin{pmatrix}
B_{1}^{\bar{\sigma}\bar{\sigma}^{\prime}}
(\scriptstyle 21 \displaystyle)&
B_{2}^{\bar{\sigma}\bar{\sigma}^{\prime}}
(\scriptstyle 21 \displaystyle)\\
-B_{1}^{\bar{\sigma}\bar{\sigma}^{\prime}}
(\scriptstyle 21 \displaystyle)&
-B_{2}^{\bar{\sigma}\bar{\sigma}^{\prime}}
(\scriptstyle 21 \displaystyle)
\end{pmatrix},
\tag{B.6}
\label{eq:B.6}
\end{equation}
where
\begin{equation}
\left.
\begin{split}
&B_{1}^{\sigma\sigma^{\prime}}
(\scriptstyle 21 \displaystyle)=
(\hat{t}G_{11}^{\sigma})
(\scriptstyle 21 \displaystyle)
G_{11}^{\sigma^{\prime}}
(\scriptstyle 12 \displaystyle)-
G_{22}^{\sigma}
(\scriptstyle 21 \displaystyle)
(G_{22}^{\sigma^{\prime}}\hat{t})
(\scriptstyle 12 \displaystyle)
\\
&B_{2}^{\sigma\sigma^{\prime}}
(\scriptstyle 21 \displaystyle)=
(\hat{t}G_{22}^{\sigma})
(\scriptstyle 21 \displaystyle)
G_{22}^{\sigma^{\prime}}
(\scriptstyle 12 \displaystyle)-
G_{11}^{\sigma}
(\scriptstyle 21 \displaystyle)
(G_{11}^{\sigma^{\prime}}\hat{t})
(\scriptstyle 12 \displaystyle)
\end{split}
\right\}.
\tag{B.7}
\label{eq:B.7}
\end{equation}

In a similar way we calculate the terminal part $\Lambda$ of the
electronic GF. Firstly we write down the expression for a matrix
element $\Pi^{11}\equiv\Lambda$ from equation (\ref{eq:3.29}).
This element is coupled with the off-diagonal element $\Pi^{21}$.
We have the following pair of coupled equations
\begin{align}
&\Lambda (\scriptstyle\textbf{12}\displaystyle) \tag{B.8}
\label{eq:B.8}
\\
&= (\hat{A}^{11}\Phi) (\scriptstyle\textbf{12}\displaystyle)
+\left[ (\hat{t}L^{11})
(\scriptstyle\textbf{4}^{\prime}\textbf{3}^{\prime}\displaystyle)
\hat{A}^{11}
(\scriptstyle\textbf{1}\textbf{4}^{\prime}\displaystyle)-
(\hat{\tilde{t}}L^{21})
(\scriptstyle\textbf{4}^{\prime}\textbf{3}^{\prime}\displaystyle)
\hat{A}^{12}
(\scriptstyle\textbf{1}\textbf{4}^{\prime}\displaystyle)
\right]\Lambda
(\scriptstyle\textbf{3}^{\prime}\textbf{2}\displaystyle)
\nonumber\\
&+\left[
(\hat{t}L^{12})
(\scriptstyle\textbf{4}^{\prime}\textbf{3}^{\prime}\displaystyle)
\hat{A}^{11}
(\scriptstyle\textbf{1}\textbf{4}^{\prime}\displaystyle)-
(\hat{\tilde{t}}L^{22})
(\scriptstyle\textbf{4}^{\prime}\textbf{3}^{\prime}\displaystyle)
\hat{A}^{12}
(\scriptstyle\textbf{1}\textbf{4}^{\prime}\displaystyle)
\right]\Pi^{21}
(\scriptstyle\textbf{3}^{\prime}\textbf{2}\displaystyle),
\nonumber
\end{align}

\begin{align}
&\Pi^{21} (\scriptstyle\textbf{12}\displaystyle) \tag{B.9}
\label{eq:B.9}
\\
&= (\hat{A}^{21}\Phi) (\scriptstyle\textbf{12}\displaystyle)
+\left[ (\hat{t}L^{11})
(\scriptstyle\textbf{4}^{\prime}\textbf{3}^{\prime}\displaystyle)
\hat{A}^{21}
(\scriptstyle\textbf{1}\textbf{4}^{\prime}\displaystyle)-
(\hat{\tilde{t}}L^{21})
(\scriptstyle\textbf{4}^{\prime}\textbf{3}^{\prime}\displaystyle)
\hat{A}^{22}
(\scriptstyle\textbf{1}\textbf{4}^{\prime}\displaystyle)
\right]\Lambda
(\scriptstyle\textbf{3}^{\prime}\textbf{2}\displaystyle)
\nonumber\\
&+\left[
(\hat{t}L^{12})
(\scriptstyle\textbf{4}^{\prime}\textbf{3}^{\prime}\displaystyle)
\hat{A}^{21}
(\scriptstyle\textbf{1}\textbf{4}^{\prime}\displaystyle)-
(\hat{\tilde{t}}L^{22})
(\scriptstyle\textbf{4}^{\prime}\textbf{3}^{\prime}\displaystyle)
\hat{A}^{22}
(\scriptstyle\textbf{1}\textbf{4}^{\prime}\displaystyle)
\right]\Pi^{21}
(\scriptstyle\textbf{3}^{\prime}\textbf{2}\displaystyle).
\nonumber
\end{align}
From here we find the zeroth order expressions for $\Pi^{11}$ and
$\Pi^{21}$
\begin{equation}
\Lambda_{0}
(\scriptstyle 1 \displaystyle \sigma_{1},
\scriptstyle 2 \displaystyle \sigma_{2})=
\delta_{12}
\bigl(
\hat{a}_{1}(\sigma_{1}\sigma_{2})\Phi
\bigr),
\tag{B.10}
\label{eq:B.10}
\end{equation}
\begin{equation}
\Pi_{0}^{21}
(\scriptstyle 1 \displaystyle \sigma_{1},
\scriptstyle 2 \displaystyle \sigma_{2})=
-\delta_{12}
\sigma_{1}(i\tau^{y})
\frac{\displaystyle \delta\Phi}
{\displaystyle \delta v_{1}^{20}}.
\tag{B.11}
\label{eq:B.11}
\end{equation}
Substituting these expressions in equations (\ref{eq:B.8}) and
(\ref{eq:B.9}) leads to the first order correction for the
terminal part:
\begin{align}
\Lambda_{1} (\scriptstyle 1\displaystyle \sigma, \scriptstyle
2\displaystyle \sigma)=& (\tau^{z}\hat{t}G\tau^{z}) (\scriptstyle
1\displaystyle \sigma, \scriptstyle 2\displaystyle \sigma) \langle
Tn_{2}^{\bar{\sigma}}n_{1}^{\bar{\sigma}} \rangle_{c} \tag{B.12}
\label{eq:B.12}
\\
+&(\tau^{z}\hat{t}G\tau^{z}) (\scriptstyle 1\displaystyle
\bar{\sigma}, \scriptstyle 2\displaystyle \bar{\sigma}) \langle
TX_{2}^{\sigma\bar{\sigma}}X_{1}^{\bar{\sigma}\sigma}
\rangle\nonumber
\\
+&(i\tau^{y}\hat{\tilde{t}}\widetilde{G}i\tau^{y})
(\scriptstyle 1\displaystyle \bar{\sigma},
\scriptstyle 2\displaystyle \bar{\sigma})
\langle TX_{2}^{02}X_{1}^{20}
\rangle\nonumber,
\end{align}
which includes bosonic GFs determined by definitions
(\ref{eq:2.20}) -- (\ref{eq:2.22}).

Fourier transformations of expressions (\ref{eq:B.4}),
(\ref{eq:B.6}), (\ref{eq:B.10}) and (\ref{eq:B.12})
give the final results for the self-energy and the
terminal part of the electronic GF. They are given
by formulas (\ref{eq:4.10}),(\ref{eq:4.12}),
(\ref{eq:4.5}) and (\ref{eq:4.7}), respectively.



\clearpage

\begin{figure}
\includegraphics[width=150mm]{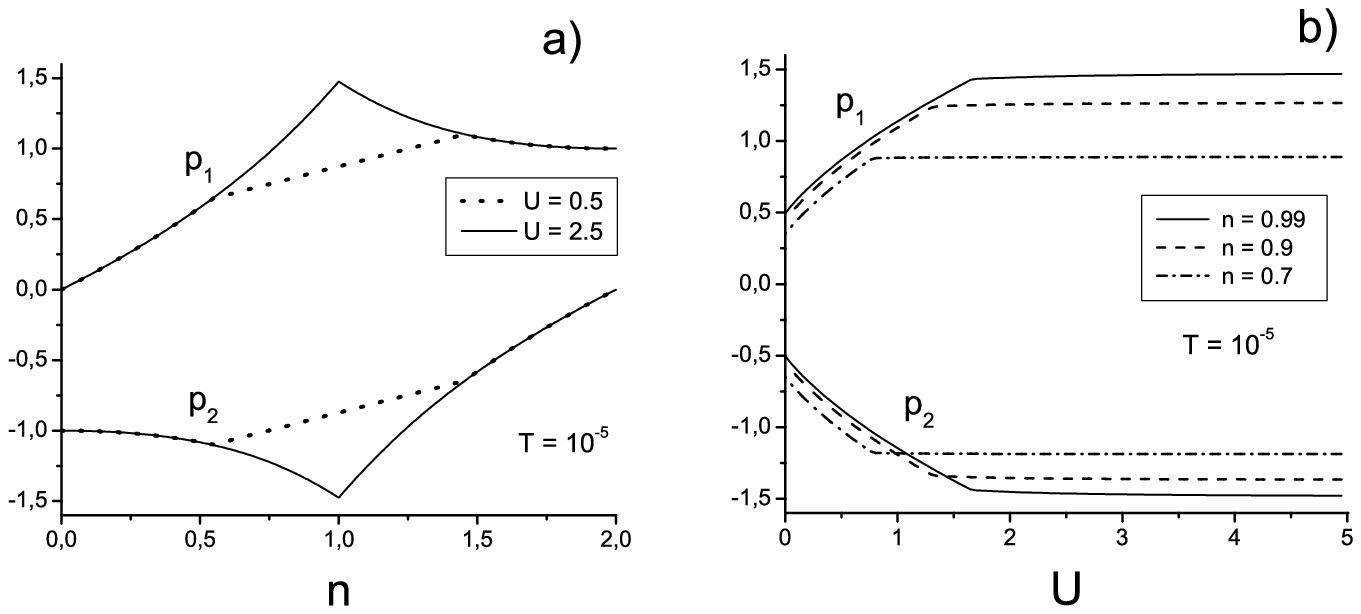}
\caption{\label{fig1_2}Dependence of parameters $p_{1}$ and $p_{2}$ on (a)
electron concentration $n$ and (b) Coulomb interaction $U$.}
\end{figure}

\clearpage

\begin{figure}
\includegraphics[width=150mm]{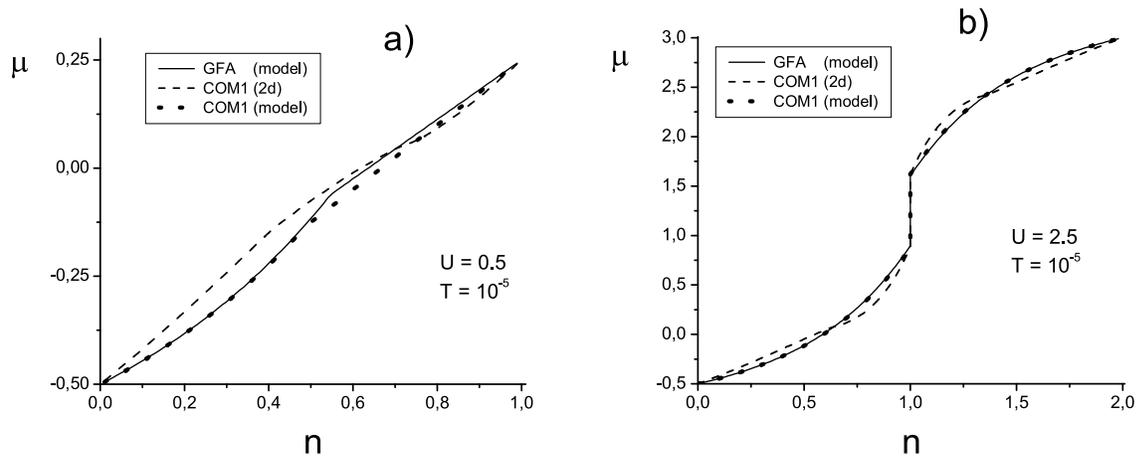}
\caption{\label{fig2_2}Chemical potential $\mu$ as a function of
electron concentration $n$ for two intervals
(a) $0<n<1$ and (b) $0<n<2$.}
\end{figure}

\clearpage

\begin{figure}
\includegraphics[width=150mm]{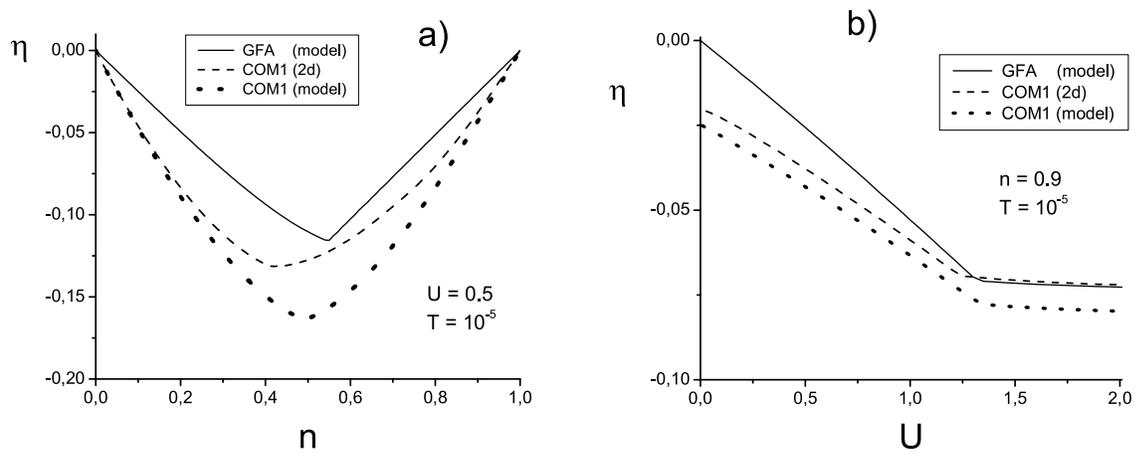}
\caption{\label{fig3_2}Parameter $\eta$ as function of $n$ and $U$.}
\end{figure}

\clearpage

\begin{figure}
\includegraphics[width=150mm]{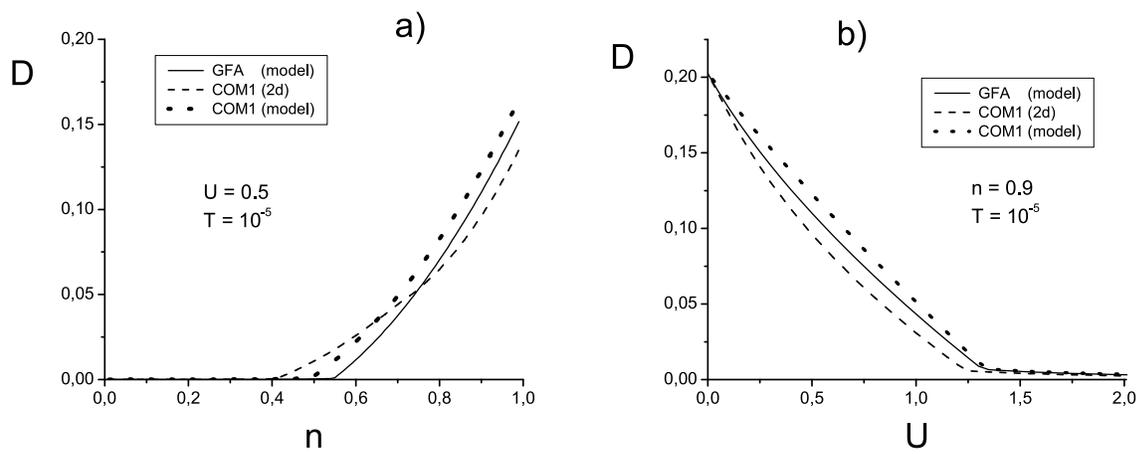}
\caption{\label{fig4_2}Parameter $D=\langle n^{\sigma}n^{\bar{\sigma}}\rangle$ of
double occupation depending on $n$ and $U$.}
\end{figure}

\clearpage

\begin{figure}
\includegraphics[width=120mm]{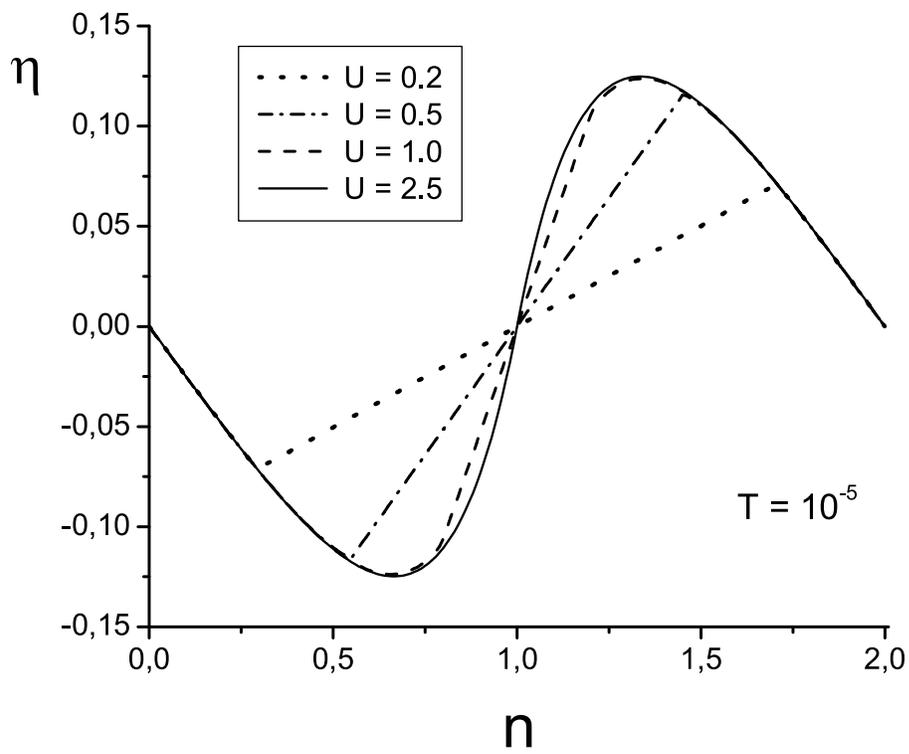}
\caption{\label{fig5_2}Dependence of parameter $\eta$ on $n$ at different values
of $U$.}
\end{figure}

\clearpage

\begin{figure}
\includegraphics[width=150mm]{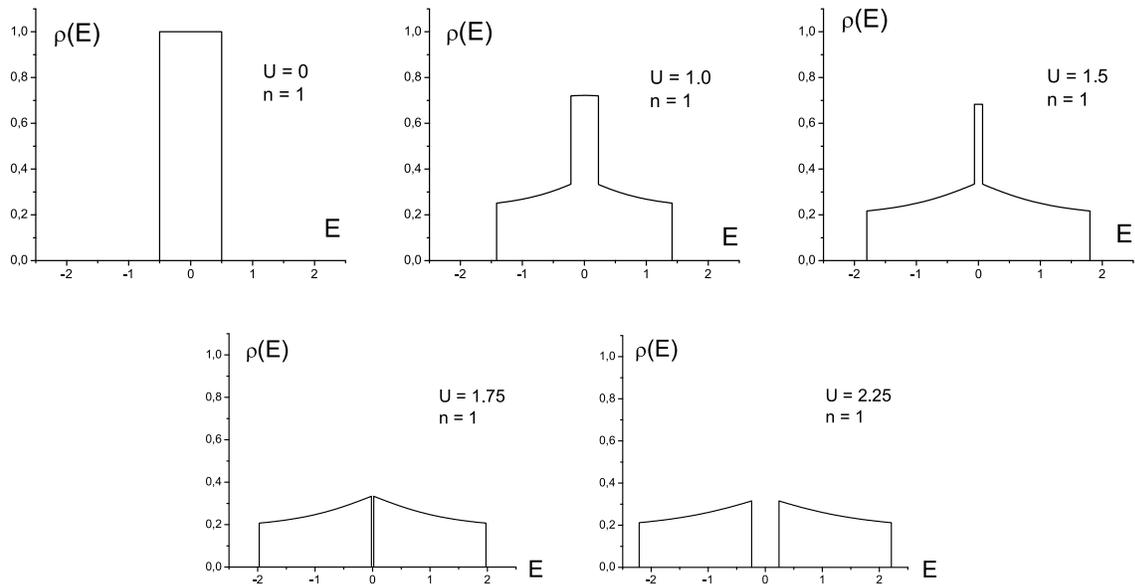}
\caption{\label{fig6_2}Evolution of the quasiparticle density of states at half
filling depending on $U$ for a model density of states
(\ref{eq:5.24}) in the bare band.}
\end{figure}

\clearpage

\begin{figure}
\includegraphics[width=120mm]{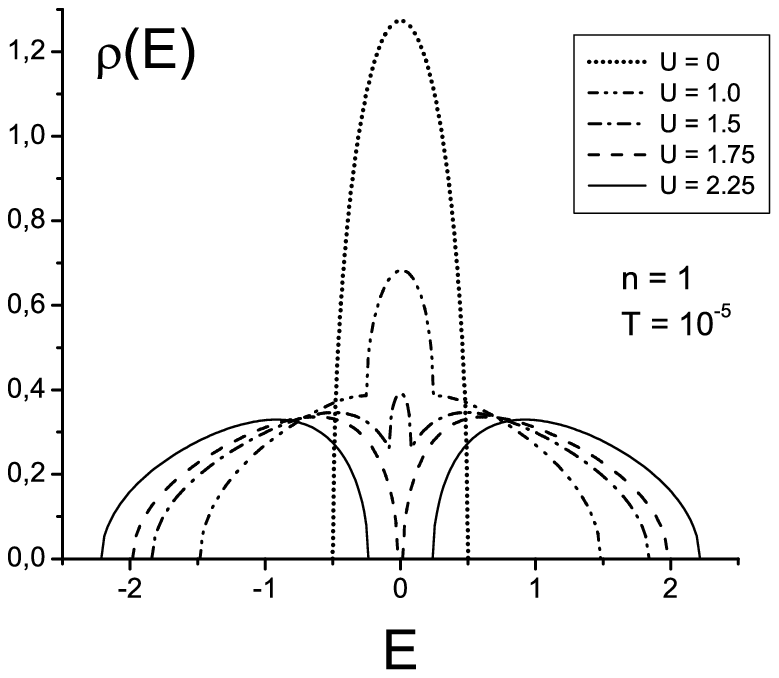}
\caption{\label{fig7_2}The same as in Fig.6 but for a semielliptic density of
states (\ref{eq:5.25}).}
\end{figure}


\begin{thebibliography}{99}

\bibitem{hu1} J. Hubbard, Proc. Roy. A., \textbf{276}, 238 (1963)

\bibitem{hu2} J. Hubbard, Proc. Roy. A., \textbf{281}, 401 (1963)

\bibitem{hu3} J. Hubbard, Proc. Roy. A., \textbf{285}, 542 (1965)

\bibitem{ro} L. Roth, Phys. Rev. \textbf{184}, 451 (1969);
\textbf{186}, 428 (1969)

\bibitem{go} E.G. Goryachev, E.V. Kuzmin, S.G. Ovchinnikov, J.
Phys. C \textbf{15}, 1481 (1982)

\bibitem{pla} N.M. Plakida, V.Yu. Yushankhai, I.V. Stasyuk, Physica
C \textbf{160}, 80 (1089)

\bibitem{man} F. Mancini, S. Marra, H. Matsumoto, Physica C
\textbf{244}, 49 (1995)

\bibitem{ave1} A. Avella, F. Mancini, D. Villani, L. Siurakshina,
V.Yu. Yushankhai, Int. J. Mod. Phys. \textbf{12}, 81 (1998)

\bibitem{ave2} A. Avella, F. Mancini, R. M\"{u}nzner, Phys. Rev. B,
\textbf{63}, 245117 (2001)

\bibitem{man2} F. Mancini and A. Avella, Eur. Phys. J. B \textbf{36}, 37 (2003)

\bibitem{ish} S. Ishihara, H. Matsumoto, S. Odashima, M. Tachiki, F.
Mancini, Phys. Rev. B \textbf{49}, 1350 (1994); H. Matsumoto, T.
Saikawa, F. Mancini, Phys. Rev. B \textbf{54}5, 14445 (1996); D.
Villani, E. Lange, A. Avella, G. Kotliar, Phys. Rev. Lett.
\textbf{85}, 804 (2000); V. Fiorentino, F. Mancini, E. Zasinas, A.
Barabanov; Phys. Rev. B \textbf{64}, 214515 (2001);  F. Mancini,
N. Perkins, N. Plakida, Phys. Lett. A \textbf{284}, 286 (2001); A.
Avella, F. Mancini, D. Villani, H. Matsumoto, Eur. Phys. J. B
\textbf{20}, 303 (2001); M. Bak, F. Mancini, Physica B
\textbf{312}, 732 (2002); A. Avella, F. Mancini, R. Hayn, Eur.
Phys. J. B \textbf{37}, 465 (2004)

\bibitem{no} W. Nolting, W. Borguel, Phys. Rev. B, \textbf{39},
6962 (1989)

\bibitem{her} T. Hermann, W. Nolting, J. Magn. Mater.
\textbf{170}, 253 (1997)

\bibitem{gutz} M.C. Gutzwiller, Phys. Rev. Lett. \textbf{10}, 159
(1963)

\bibitem{bar} S.E. Barnes J. Phys. F. \textbf{6}, 1375 (1976);
\textbf{7}, 2673 (1977)

\bibitem{kot} G. Kotliar, A.E. Ruckenstein, Phys. Rev. Lett.
\textbf{57}, 1362 (1986)

\bibitem{lil} L. Lilly, A. Muramatsu, W. Hanke, Phys. Rev. Lett.
\textbf{65}, 1379 (1990)

\bibitem{fre} R. Fresard, M. Dzierzawa, P. W\"{o}lfle Europhys.
Lett. \textbf{65}, 1379 (1990)

\bibitem{metz} W. Metzner, D. Vollhardt, Phys. Rev. Lett.
\textbf{62}, 324 (1989)

\bibitem{geo1} A. Georges, G. Kotliar, Phys. Rev. B \textbf{45},
6479 (1992)

\bibitem{geo2} A. Georges, G. Kotliar, W. Krauth, M.J. Rozenberg,
Rev. Mod. Phys. \textbf{68}, 13 (1996)

\bibitem{ba} G. Baym, L.P. Kadanoff, Phys. Rev. \textbf{124}, 287
(1961)

\bibitem{ka} L.P. Kadanoff, G. Baym, Quantum Statistical
Mechanics, Benjamin, New York, 1962

\bibitem{ru} A.E. Ruckenstein, S. Schmitt-Rink, Phys. Rev. B
\textbf{38}, 7188 (1988)

\bibitem{ku} M.L. Kulic, R. Zeyher, Phys. Rev. B \textbf{49},
4395 (1994)

\bibitem{ze} R. Zeyher, M.L. Kulic, Phys. Rev. B \textbf{51},
1234 (1995)

\bibitem{iz1} Yu.A. Izyumov, N.I. Chaschin, The Physics of Metals
and Metallography, \textbf{92}, N5, 451 (2001); \textbf{92}, N6,
531 (2001); \textbf{93}, N1, 18 (2002); \textbf{94}, N6, 527
(2002); \textbf{94}, N6, 539 (2002);


\bibitem{plak1} N.M.Plakida, Physica C \textbf{282--287}, 1737 (1997)   

\bibitem{plak2} N.M.Plakida, L.Anton, S.Adam and Gh.Adam, JETP \textbf{97}, 331 (2003)

\bibitem{ave3} A. Avella, S. Krivenko, F. Mancini, N.M. Plakida, J. Magn. Magn. Mater. \textbf{272}, 456
(2004)

\bibitem{ave4} A. Avella, F. Mancini, V. Turkowski; Phys. Rev. B \textbf{67}, 115123 (2003)

\bibitem{iz2} Yu.A. Izyumov, N.I. Chaschin, V.Yu. Yushankhai,
Phys. Rev. B \textbf{65}, 214425 (2002)

\bibitem{iz3} Yu.A. Izyumov, Yu.N. Skryabin, Basic Models in
Quantum Theory of Magnetism, Ural Division of the Russian Academy
of Sciences, Ekaterinburg, 2002 (in Russian)

\bibitem{iz4} Yu. A. Izyumov, in Lectures on the Physics of Highly
Correlated Electron Systems VII, Seventh Training Course in the
Physics of Correlated Electron Systems and High-Tc
Superconductors, ed. A. Avella, F. Mancini, AIP Conference
Proceedings 678, 2003

\bibitem{har} A. Harris, R. Lange, Phys. Rev. \textbf{157}, 295
(1967)

\bibitem{fors} D. Forster; Hydrodynamical Fluctuations,
Broken Symmetry and Correlation Functions, New York:
Benjamin, 1975

\bibitem{ya} C. N. Yang; Phys. Rev. Lett \textbf{63}, 2144
(1989)

\bibitem{shen} S. Q. Shen , X. C. Xie, Condens Matter, \textbf{8}, 4805
(1996)

\bibitem{bruks} H. Bruks, J. C. A. d'Auriac, Phys. Rev. B \textbf{55}, 9142
(1997)

\bibitem{ave6} A. Avella, F. Mancini, D.Villani; Cond-mat /9807402

\bibitem{juk} G. Jakely, N.M. Plakida; Theor. and Mat. Phys. (Russian)
\textbf{114}, 3, 426
(1998)

\bibitem{mori} H. Mori; Progr. Theor. Phys. \textbf{33}, 423 (1965);
\textbf{34}, 426
(1965)

\bibitem{he} J.A. Hertz, D.M. Edwards, J. Phys. F. \textbf{3},
2174 (1973); D.M. Edwards, J.A. Hertz, J. Phys. F. \textbf{3},
2191 (1973)


\bibitem{zar} A.V. Zarubin, V.Yu. Irkhin, Fiz. Tverd. Tela.
\textbf{41}, 1057 (1999); (Phys. Solid State \textbf{41}, 963
(1999))

\end{thebibliography}
\end{document}